\documentclass[12pts,fleqn,usenatbib,useAMS]{mnras}

\usepackage{amsmath}
\usepackage{color}
\usepackage{subfigure}
\usepackage{float}
\usepackage{natbib}
\usepackage[T1]{fontenc}

\RequirePackage{fixltx2e}

\def\la{\mathrel{\hbox{\rlap{\hbox{\lower4pt\hbox{$\sim$}}}\hbox{$<$}}}}

\def\ga{\mathrel{\hbox{\rlap{\hbox{\lower4pt\hbox{$\sim$}}}\hbox{$>$}}}}

\usepackage{journal_names}
\usepackage{color}

\usepackage{url}

\newcommand\farcss{\mbox{$.\!\!\!^{\prime\prime}$}}
\def\farcm{\mbox{.\kern -0.5ex\raisebox{.6ex}{\scriptsize$\prime$}}}

\def\farcss{
 \mbox{ 
  \kern  0.13ex. 
   \kern -0.95ex\raisebox{.6ex}{\scriptsize$\prime\prime$}
  \kern -0.1ex
 }
}

\usepackage{graphicx,times}

\usepackage{float}

\title[Pebble-based planet formation]{Probing the impact of varied migration and gas accretion rates for the formation of giant planets in the pebble accretion scenario}

\author[Ndugu et al.]{
N.Ndugu,$^{1}$\thanks{E-mail: nndugu@must.ac.ug}
B.Bitsch,$^{2}$\thanks{E-mail: bitsch@mpia.de}
A. Morbidelli,$^{3}$
A. Crida,$^{3,4}$
E.Jurua$^{1}$\thanks{E-mail: ejurua@must.ac.ug}
\\
$^{1}$Department of Physics, Mbarara University of Science and Technology, 
Mbarara, \textsc{Uganda}\\
$^{2}$Max-Planck-Institut for Astronomy, K{\"o}nigstuhl 17, D-69117 Heidelberg, 
\textsc{Germany}\\
$^{3}$Observatoire de la C\^ote d'Azur / CNRS, Laboratoire Lagrange, Boulevard 
de l’Observatoire, CS 34229, 06300 Nice, \textsc{France}\\
$^{4}$Institut Universitaire de France, 103 Boulevard Saint-Michel, 75005 
Paris, \textsc{France}
}

\pubyear{2020}

\begin{document}
\label{firstpage}
\pagerange{\pageref{firstpage}--\pageref{lastpage}}
\maketitle

\begin{abstract}
The final orbital position of growing planets is determined by their migration speed, which is essentially set by the planetary mass. Small mass planets migrate in type~I migration, while more massive planets migrate in type~II migration, which is thought to depend mostly on the viscous evolution rate of the disc. A planet is most vulnerable to inward migration before it reaches type~II migration and can lose a significant fraction of its semi-major axis at this stage. We investigated the influence of different disc viscosities, the dynamical torque and gas accretion from within the horseshoe region as mechanisms for slowing down planet migration. Our study confirms that planets growing in low viscosity environments migrate less, due to the earlier gap opening and slower type~II migration rate. We find that taking the gas accretion from the horseshoe region into account allows an earlier gap opening and this results in less inward migration of growing planets. Furthermore, this effect increases the planetary mass compared to simulations that do not take the effect of gas accretion from the horseshoe region. Moreover, combining the effect of the dynamical torque with the effect of gas accretion from the horseshoe region, significantly slows down inward migration. Taking these effects into account could allow the formation of cold Jupiters (a~$>$~1~au) closer to the water ice line region compared to previous simulations that did not take these effects into account. We thus conclude that gas accretion from within the horseshoe region and the dynamical torque play crucial roles in shaping planetary systems.

\end{abstract}
%

\section{Introduction} \label{Introduction}

The number of observed extrasolar planets has tremendously grown over the 
last decades, since the discovery of the first exoplanet around a main sequence 
star \citep{Mayor1995}. These observed extrasolar planets are very diverse in 
their properties (e.g. mass, semi-major axis, eccentricity, etc). In particular, it 
is now clear that the hot jupiters are a minority, and that most giant planets 
orbit a few aus from their host star \citep{Howard2010}. Nevertheless, giant 
planets are the minority, instead the population of planets is dominated by 
close-in hot super-Earths \citep{Fressin2013} and by ice giants orbiting around 
the ice-line \citep{Cassan2012,Suzuki2016}. Matching all of these different populations of 
planets is a challenge for planet formation theories.

Studies of giant exoplanet occurrences show that giant planets are seen 
more frequently around high metallicity and massive stars 
\citep{Santos2004,Fischer2005,Johnson2010} which is in agreement with core 
accretion paradigms \citep{Pollack1996}. Recently, a gas giant has been identified orbiting a low 
mass star \citep{Morales2019}, presenting a challenge to planetary growth and 
migration models, but in support of gravitational instability model \citep{Nayakshin2015MNRAS.454...64N}. Therefore 
there is still no clear global understanding to explain the rich diversity of 
the observed extrasolar planets.

The first attempts to explain the distribution of exoplanets have been
done in the so called population synthesis studies that started about a
decade ago \citep{Ida2004}. Planet population
synthesis models have become progressively more complex over the
years \citep{Alibert2005,Ida2008a,Ida2008b,Mordasini2009,Miguel2011,Alibert2011,Ida2013,Fortier2013,Dittkrist2014,Ronco2017,
BitschJohansen2017,Ndugu2018,Chambers2018,Brugger2018, Ida2018,Johansen2019,Ndugu2019,Cridland2019}. Most of these planet population synthesis simulations model the planetary growth phase due to the accretion of planetesimals \citep[e.g.][]{Alibert2005,Ida2008a,Ida2008b,Mordasini2009,Miguel2011,Alibert2011,Ida2013,Fortier2013,Dittkrist2014,Ronco2017}. However, the accretion of planetesimals is inefficient if most of the mass is carried by planetesimals larger than a few 10 km in size \citep[e.g.][]{TanakaIda1999,Thommes2003,Levison2010,JohansenBitsch2019}. Most of the planetesimal-based planet population synthesis studies thus make use of small planetesimals (below 1 km in size) to achieve a fast enough growth of planets \citep[e.g.][]{Mordasini2009}, even though there is no evidence in the solar system that planetesimals were that small in the inner regions \citep{Bottke2005,Morbidelli2009,Singer2019}.

A fresh attempt to model planet population synthesis via pebble accretion appeared in the recent years 
\citep[e.g.][]{AliDib2017,BitschJohansen2017,Ndugu2018,Chambers2018,Brugger2018,Ida2018,Johansen2019,Ndugu2019}. In these planet population syntheses models, solid accretion is solely from 
accretion of mm-cm sized particles \citep{OrmelKlahr2010,JohansenLarcda2010,Lambrechts2012,Lambrechts2014}. These 
models include type~I \citep{Paardekooper2011} and type-II migration 
\citep{Baruteau2014}, as well as gas accretion \citep{Pisso2014,Machida2010}. 
\cite{Ndugu2018, Ndugu2019} follow the disc model of  \cite{Bitsch}. 
\cite{Johansen2019} used a modified pebble-accretion approach with relatively 
smaller pebble sizes than used in previous studies
\citep{BitschJohansen2017,Ndugu2018,Chambers2018,Brugger2018,Ndugu2019}, showing how planetary growth can out perform planet migration. The pebble-based planet population synthesis models compute the final mass and location of planets depending on the initial time when the planetary embryo was 
placed in the disc, the location at which they started to grow and the disc 
parameters.

Most of the models preferentially reproduce some of the subsets of the observed exoplanets, but not the full statistics of the observed extrasolar planets. The planet population synthesis models mainly suffer from a mismatch between the migration and the growth speed.  \cite{Johansen2019} found that planets could be saved from being lost to the host star via radial migration by using smaller pebble size and using slower Type~II migration following the simulations of \cite{Kanagawa2018}.

\citet[][B15, hereafter]{Bitsch-etal-2015} used a pebble based planet formation model with a high viscosity disc and found a quite surprising result. They find that for a Jupiter-mass planet to be at a few au from the central star at the end of the protoplanetary disc lifetime, its seed has to form beyond 20~au, towards the end of the disc's evolution (i.e. at $t=1.5-2$~Myr, for a disc's lifetime of 3~Myr). However, the most favorable place for giant planet formation is presumably the water ice line region ( at most $\sim 7$~au from the central star in a young disc; \cite{Bitsch}), where planetesimals can form rapidly due to the local pile-up of pebbles at this location \citep{IdaGuillot2016,Schoonenberg2017,Drazkowska2017A&A...608A..92D} and where the subsequent pebble-accretion process is the most efficient \citep{Morbidelli2015}. How could the seeds of the giant planets form beyond 20~au and not at the water ice line? A possibility is that all planets forming within 20~au or earlier than $\sim 2/3$ of the disc's lifetime ended by migration into the Sun. But there is no evidence in the Solar System for the migration of massive planets through the terrestrial planet region. Moreover, it is questionable whether planets can be engulfed by the parent star. The discovery of many systems of close-in super-Earths argues that protoplanetary discs have inner edges, where planet migration is stopped \citep{Masset2006,Ogihara2015,Flock2019}. Thus, it is difficult to accept the idea that many planets formed in the Solar System but disappeared into the Sun, and that the present giant planets are just the "last of the Mohicans" \citep{Lin1997,Laughlin1997}.

The reason why in the model of B15 the giant planets have to start forming far and late is that planet migration, at face values,
is too fast. Unlike in earlier works, type-I migration of solid cores does not appear to be the critical phenomenon. This is because, in the pebble accretion scenario, the solid cores of planets grow very quickly, and therefore they don't have the time to migrate far away from their initial location. Moreover, there are in the disc some outward migration zones \citep{Paardekooper2011,Bitsch2014,Bitsch}, that block inward type~I migration for moderate-mass planets. Instead, the most
dangerous phases in terms of migration occur when the planet is already very massive. The first critical phase occurs when the planet exceeds the so-called {\it pebble isolation mass} \citep{Lambrechts2014b}. When the planet exceeds a mass of the order of 20~Earth masses (this value scales as $ (H/r/0.05)^3$, where $H/r$ is the local
scale-height of the disc), the planet stops accreting pebbles because the latter remain blocked at the pressure bump generated outside of the planet's orbit by the the opening of a shallow gap in the disc. Hence, the accretion of the planet slows down considerably, because only gas accretion remains possible. Inward migration is fast in this phase. In fact, the entropy-driven
corotation torque that drives outward migration of medium-mass planets \citep{Paardekooper2011} is not operational for planets of tens of Earth masses, because the corotation torque becomes saturated, particularly in late discs \citep{Bitsch2014,Bitsch}. 

The second critical phase is that of type-II migration, occurring when the planet becomes massive enough to open a deep gap in the disc. Although type-II migration occurs at the viscous accretion rate of the disc, for realistic rates consistent with observations \citep{Hartmann1998, Manara2012} type~II migration is fast enough to move giant planets by several aus on a Myr
timescale. \cite{Nelson2000} also identified type-II migration to be mostly responsible for bringing giant planets too
close to the star compared to observations.

Recently, \citet[][CB, hereafter]{Crida2017} have shown that a planet accreting gas at a high enough rate can open a gap in the disc before its mass exceeds the nominal value for gap opening through planetary torques \citep{Crida2006}. This may 
allow the planet to transition to Type~II migration at lower masses than envisioned in B15. In this phase, the planet accretes gas mostly from its co-orbital region and therefore it can grow faster than the rate at which the disc supplies mass through radial transport, until the horseshoe region is depleted, which then marks the transition to type~II migration.

Another important result originating from recent studies \citep{Nelson2013,Stoll2014} is the fact that discs may have a very small mid plane viscosity ($\alpha$ of the order of $ 10^{-4}$ or less, where $\alpha$ is the parameter governing the viscosity in the \cite{Shakura1973} prescription) because the Magneto-Rotational Instability (MRI) is quenched at all scale-heights in the disc \citep{Turner2014}. 

The goal of this paper is thus to investigate the potential impact of lower disc viscosity, gap opening by gas accretion \citep{Crida2017,Bergez2020arXiv201000485B} and the influence of dynamical torques \citep{Paardekooper2014} on the formation of planets. The target is to find in the disc where and when the dynamical torque and gap opening by gas accretion becomes important.

Planet population synthesis simulations should aim to explain all the data available from exoplanet observations at once. This requires not only to include all processes relevant for planet formation (e.g. disc evolution, solid and gas accretion, planet migration), but also to vary the initial conditions within each simulation (e.g. alpha viscosity parameter, dust-to-gas ratio, disc mass and size, embryo starting time and position). In our study, we focused to understand the impact of a new recipe for gas accretion and for planet migration on planet formation via global planetary evolution map approach of \cite{Bitsch-etal-2015}. We thus limited ourselves to one simple disc model and a controlled change in the alpha viscosity parameter to understand when and how our new recipes for gas accretion and planet migration affect the picture of planet formation, which might not be possible if all parameters are drawn at random as done in comprehensive planet population synthesis models.

This paper is organized as follows. In section~\ref{method}, we highlight the disc model and planet formation model used. In section~\ref{Gapopening}, we introduce the concept of gap opening by accretion, following CB. We additionally introduce a prescription for the dynamical torques in the last part of section~\ref{Gapopening} with the goal of reducing the fast inward type-I migration in low viscosity environments. All the results will be presented showing individual evolution of planets in terms of their mass and semi major axis, as well as global maps like those of B15, showing the final location and mass of planets as a function of the time and the location at which they started to grow. In section~\ref{when}, we explore where and when the CB formalism and the dynamical torque becomes effective in gas giant planet formation. Section~\ref{Conclusions} summarizes the results, putting them into perspectives for future investigations.

\begin{table}
\caption{Model descriptions used in the planet formation simulations. The acronym NN, CB and Dyn describes the models of \citet{Ndugu2018}, \citet{Crida2017} and \citet{Paardekooper2014}, respectively, where \citet{Paardekooper2014} describes the model of the dynamical torques. $\alpha$ is the viscosity used in the simulation. The rest of the symbols of the model are derived from a combination of the aforementioned models. The nom abbreviation symbolizes the nominal migration and gas accretion parameter values as in \citet{Bitsch-etal-2015} and \citet{Ndugu2018}.}
\label{tab:ambiguous}
\begin{center}
\footnotesize
\begin{tabular}{lccccccccccccc}

Model & $\alpha$ &Gas accretion &Torque&\\ 
           &  &  &  &  &  \\ 
\hline
			       NN 	& 0.005	& nom & nom & \\ 
			       $\rm 5e-4$NN 	& 0.0005 	& nom & nom & \\
			       $\rm 1e-4$NN 	& 0.0001 	&  nom & nom & \\         

			       CB 	& 0.005 	& CB & nom &  \\ 
			      $\rm 1e-3$CB	& 0.001  	& CB & nom &  \\ 
			      $\rm 5e-4$CB	& 0.0005  	&CB & nom &  \\  
			      $\rm 1e-4$CB	& 0.0001  	&  CB & nom & \\

                   NNDyn	& 0.005  	& nom & nom+Dyn &\\
                   $\rm 1e-4$NNDyn	& 0.0001  	&  nom & nom+Dyn &\\
                   
                   CBDyn	& 0.005 	& CB & nom+Dyn& \\ 
			      $\rm 1e-3$CBDyn	& 0.001 	& CB & nom+Dyn& \\
			      $\rm 5e-4$CBDyn	& 0.0005  	&  CB & nom+Dyn& \\
			      $\rm 1e-4$CBDyn	& 0.0001  	& CB & nom+Dyn &\\
			       
\end{tabular}
\end{center}
\end{table}

\section{Methods} \label{method}
\subsection{Disc model}
We start by setting up a model for the gas component of the protoplanetary
disc. By assuming that the gas accretion rate is independent of radius (steady state assumption), we define the surface density radial profile for the disc as
\begin{align}
\Sigma_{\rm g} = \beta \left( \frac{r}{\rm au} \right)^{-15/14},
\end{align}
where we set $\beta = 1500 \exp(-\frac{t}{\tau_{\rm disc}})$ in order to mimic gas dissipation, similar to the approach of \citet{McNeil_2005}, \citet{Walsh2011} and \citet{Lambrechts2014}. This radial power law profile is typical of viscous evolution of an accretion disc \citep{Lynden-Bell1974} and is in addition supported by the observed disc profile of the nearby protoplanetary disc around the star TW Hya \citep{Andrews_2012}. For simplicity we used $\tau_{\rm disc} = 3$\,Myr \citep{Haisch_2001}.
Our thermal profile of the disc ($T,c_{\rm s}, H/r$) follows the standard \cite{Chiang1997ApJ...490..368C} prescription. In particular,
\begin{equation}
H/r=h_0\left(\frac{r}{\rm au}\right)^{2/7}, 
\end{equation} where we either used $h_0=0.033~{\rm or}~h_0=0.025$ to check the influence of the disc's thermal structure on the migration rates. 

We note that throughout this work, our nominal disc model uses $h_0=0.025$. The herein adopted flared disc model removes the outward migration regions for small-mass planets due to the entropy-driven corotation torque.
Here, one should note that the disc temperature is not influenced by the disc viscosity since the \cite{Chiang1997ApJ...490..368C} prescription does not account for viscous heating. Therefore our model deviates significantly from a viscously heated model in the inner regions of the disc \citep[e.g,][]{Bitsch}. However, because in in this study we will mostly deal with low-viscosities in the disc, the \cite{Chiang1997ApJ...490..368C} temperature profile is an acceptable approximation.

\subsection{Planet formation model}\label{planetform}

Here we start by presenting similar growth tracks as used in \citet[][NN, hereafter]{Ndugu2018} but utilizing the simple flared disc  profile from \cite{Chiang1997ApJ...490..368C}  as the reference model for the rest of the paper. In our model, the planetary embryos that grow and migrate through the disc start at the pebble transition mass (when Hill accretion becomes efficient, see  \cite{Lambrechts2012}).

\begin{figure}
 \centering
 \includegraphics[width=\columnwidth]{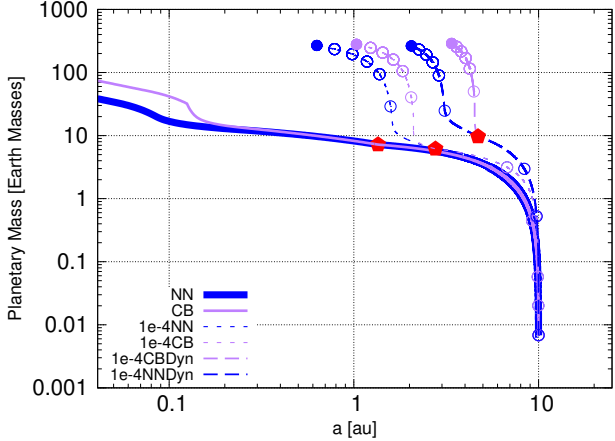}
 \caption{Evolution of planets implanted at 10~au in a disc of lifetime of 3~Myr at 0.1~Myr using the  different models  described in Table~\ref{tab:ambiguous} with the $\alpha$ parameters stated therein. The simulation was terminated when the planet reached 0.04~au or evolved for 3~Myr (marked by a solid dot). The circles on the curves are placed at $t_0, 2t_0, 3t_0,  5t_0,  10t_0,  15t_0,  20t_0,  25t_0$ and $30t_0$~Myr, respectively. The pebble isolation mass is depicted by the red filled polygonal mark. The solid blue NN track was made thick for a clear differentiation with the solid CB track which basically follows exactly the same growth trajectory.}
  \label{fig:1}
\end{figure} In this section, we study the evolution of a single planet in time. The planetary seed is introduced at time $t_0=0.1$~Myr at a distance $a_0=10$~au from the central star and in a disc with a lifetime of 3~Myr. In addition, we used uniform opacity in the planetary envelope (when they exist; $\kappa_{\rm env}=0.05~\rm cm^{2}g^{-1}$). These simulations will be a reference with comparison to the results obtained in the low viscosity disc models, discussed in subsection~\ref{consequences}. Our envelope opacity choice is in agreement with the study by \cite{Movshovitz2008}, who found that the dust grain opacity in most of the radiative zone of the planet's envelope is of the order of $10^{-2}$. This is also in agreement with \cite{Mordasini2015}. The amount of pebbles that are available to the planet in the disc is crucial in determining the final mass and distance of the planet. We therefore mimic the pebble supply rate by a pebble flux, $\dot{M}_{\rm peb}$ which we set to 
 
 \begin{equation}
 \dot{M}_{\rm peb} = 2\times 10^{-4} \exp\left(-\frac{t}{\tau_{\rm disc}}\right){\rm \frac{M_{E}}{year}}.
\end{equation} The pebble surface density at the planets position is given by
\begin{equation}\label{eq:Sigmapebb}
 \Sigma_{\rm peb}= \frac{\dot{M}_{\rm peb}}{2\pi r_{\rm p} \eta v_{\rm k} \tau_{\rm f} }
\end{equation} where $r_{\rm p}$ denotes the planet's semi-major axis. $\eta,~v_{\rm k}~{\rm and}~\tau_{\rm f}$ are the pressure support, the Keplerian speed and the Stokes number, respectively. For simplicity we adopt a uniform Stokes number, $\tau_{\rm f}=0.1$, in all our simulations\footnote{This approximation results in an increase of the physical particle sizes from 0.9 to 29 cm from the outer to the inner disc}.
  
Planetary cores initially accrete material via the inefficient 3D pebble 
accretion branch \citep{Lambrechts2012} until they become massive enough that 
their Hill radius exceeds the pebble scale height  ($r_{\rm H}>H_{\rm peb}$, 
see \cite{Morbidelli2015} for more discussion). Planetary embryos with $r_{\rm 
H}>H_{\rm peb}$ accrete pebbles in the efficient 2D Hill regime at a rate given 
by \cite{Lambrechts2012} as

\begin{equation}
  \dot{M} = 2 \left( \frac{\rm \tau_{\rm f}}{0.1} \right)^{2/3} \varOmega 
R_{\rm H}^2
  \Sigma_{\rm peb} \, .
  \label{eq:Mdot}
\end{equation} $\Omega$ is the Keplerian frequency at the location of the planetary embryo. We acknowledge that for a given particle size the Stokes number is non uniform along the disc and that a change of the Stokes number influences planet growth (see Equation~\ref{eq:Mdot}). The choice of a fixed value of the Stokes number is a simplification, also made in \cite{Johansen2019}, justified by the fact that the pebble accretion rate is parameterized by the Stokes number and we did not include a realistic model of dust size evolution in this work, as our aim is mostly to show the influence of the gas accretion rate from the horseshoe region and the effects of the dynamical torque on the accretion and migration of planet. The different choices of how the Stokes number influences our planet formation model is discussed in appendix~\ref{Stokes}. In the described solid accretion regime, planets migrate in the type~I migration regime that scales linearly with the mass of the growing embryo. The type~I migration rate is calculated using the torque prescription of \cite{Paardekooper2011}. This torque prescription consists of the Lindblad, the barotropic corotation and the entropy related corotation torque. The total torque, $ \Gamma_{\rm tot} $ acting on the planet is given by 
\begin{equation}
    \label{eq:tottorque}
   \Gamma_{\rm tot} = \Gamma_{\rm L} + \Gamma_{\rm C}.
\end{equation}
  $ \Gamma_{\rm L}$ and $ \Gamma_{\rm C}$ are the Lindblad and corotation torques, respectively. The Lindblad and corotation torques strongly dependent on the local radial gradients of gas surface density, $ \Sigma_{\rm g} \propto r^{-\lambda} $, temperature $T \propto r^{-\beta} $, and entropy $S \propto r^{-\varepsilon} $, with $ \varepsilon = {\beta} + \left(\gamma - 1.0\right)\lambda $ and $\gamma = 1.4  $ is the adiabatic index.
  
  At low $\alpha$ values, type~I migration is increasingly fast with increasing planetary mass due to early saturation of corotation torques. However, the speed of Type~I migration can be  reduced if we consider the dynamical corotation torque \citep{Paardekooper2014} or thermal torque due to heating by the accreting pebbles \citep{Benitez2015}\footnote{The maximal accretion rates for our planets are around $10^{-4}$ Earth masses/year, which is a factor of a few lower than what is needed for the heating torque to operate \citep{Baumann2020}. We therefore do not  incorporate the contribution of thermal torque in our simulations.}. The dynamical corotation torque is particularly effective in reducing type~I migration in low-viscosity discs with surface density profile shallower than $1/r^{3/2}$, like the one assumed here. We discussed the detailed implementation of dynamical corotation torque in appendix~\ref{dyntorque}. We remind the reader that type~I migration changes in inviscid disc, where the \cite{Paardekooper2011} prescription might not hold. The focus of our paper is not inviscid discs, but to probe the impact of lower migration rates via lower viscosities in discs. We acknowledge that at lower visocities, although type~II migration is slowed, type~I migration is too fast for forming large planets due to early saturation of corotational torques. We however, reduced this by invoking the dynamical corotation torque in our migration prescription.
  
We start a planetary seed at $a_{0} = 10$~au and at $t_0 =0.1$~Myr and let it 
grow and migrate. The reference growth track is depicted by the blue 
solid curve in Figure~\ref{fig:1}. The seed started at a mass corresponding to 
the transition between the Bondi regime and the Hill regime in the pebble 
accretion process \citep{Lambrechts2012} and was allowed to grow by accreting 
pebbles while exhibiting Type~I migration, until pebble isolation mass 
\citep{Lambrechts2014,Bitsch2018,Ataiee2018}. In this paper, the pebble isolation mass formula from \cite{Bitsch2018},
\begin{eqnarray}
\label{eq:Iso}
 {M_{\rm iso}}&\approx&25\left(0.34\left(\frac{\log_{10}(0.001)}{\log_{10}(\alpha)}\right)^4 + 0.66\right)\nonumber \\ &&\left(1 - \left(P_{\rm grad} + 2.5\right)/6\right)\left(\frac{H/r}{0.05}\right)^{3}{\rm M}_{\rm E}
\end{eqnarray} is used. $P_{\rm grad}$ corresponds to the local radial pressure gradient in the unperturbed protoplanetary disc.

 At pebble isolation mass (mass corresponding to the filled polygonal mark for Figure~\ref{fig:1}, in this 
case), pebble accretion is halted and the core starts accreting gas. However, 
the planet keeps evolving under pure Type~I migration until it opens a gap with 
50~\% depth. As the planet grows further, it carves a deeper gap, which will 
eventually lead it into the Type~II migration regime when the gap reaches 
90~\% depth \citep{Crida2006}. The migration rate in the intermediate 
phase between a gap depth of 50 and 90~\% is computed as a linear interpolation 
between the Type~I and Type~II migration rates, based on an analytic estimate of the depth of the gap \citep{Crida2007}. \cite{Kanagawa2018} proposed a migration formula where type~I migration is smoothly transitioned into type~II migration without interpolation as in our case. We did not adopt the \cite{Kanagawa2018} prescription for
 \begin{figure*}
        \centering
         \includegraphics[width=\columnwidth, 
height=2.45in]{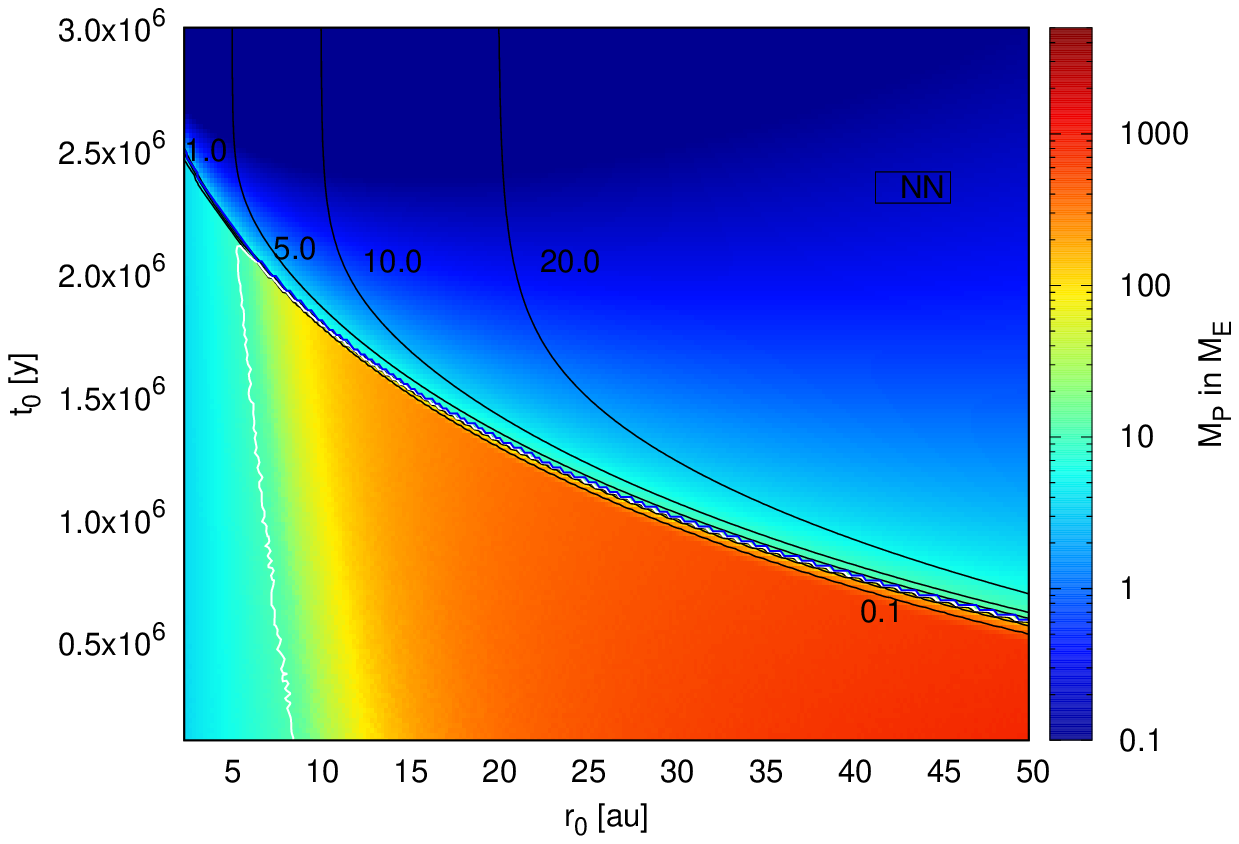}\quad
        \includegraphics[width=\columnwidth, 
height=2.45in]{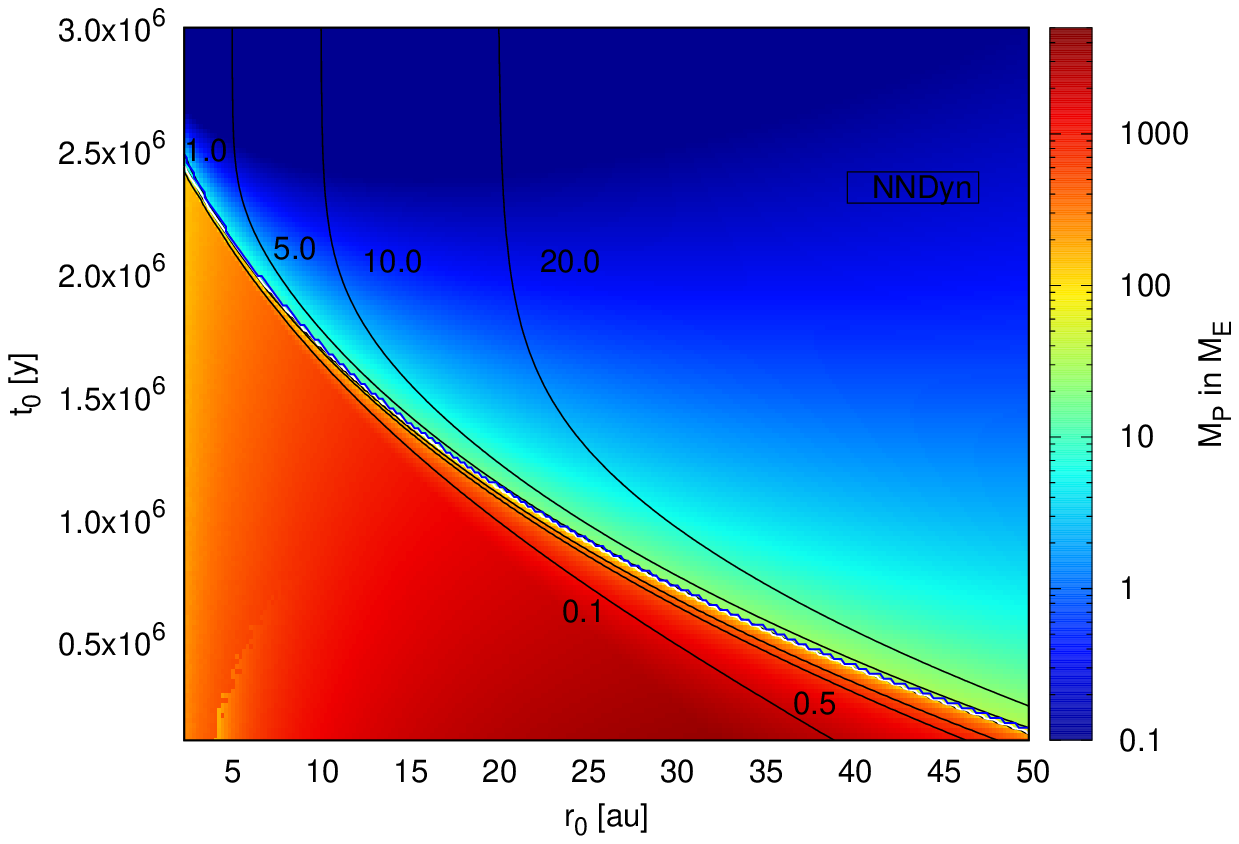} \quad
         \includegraphics[width=\columnwidth, height=2.45in]{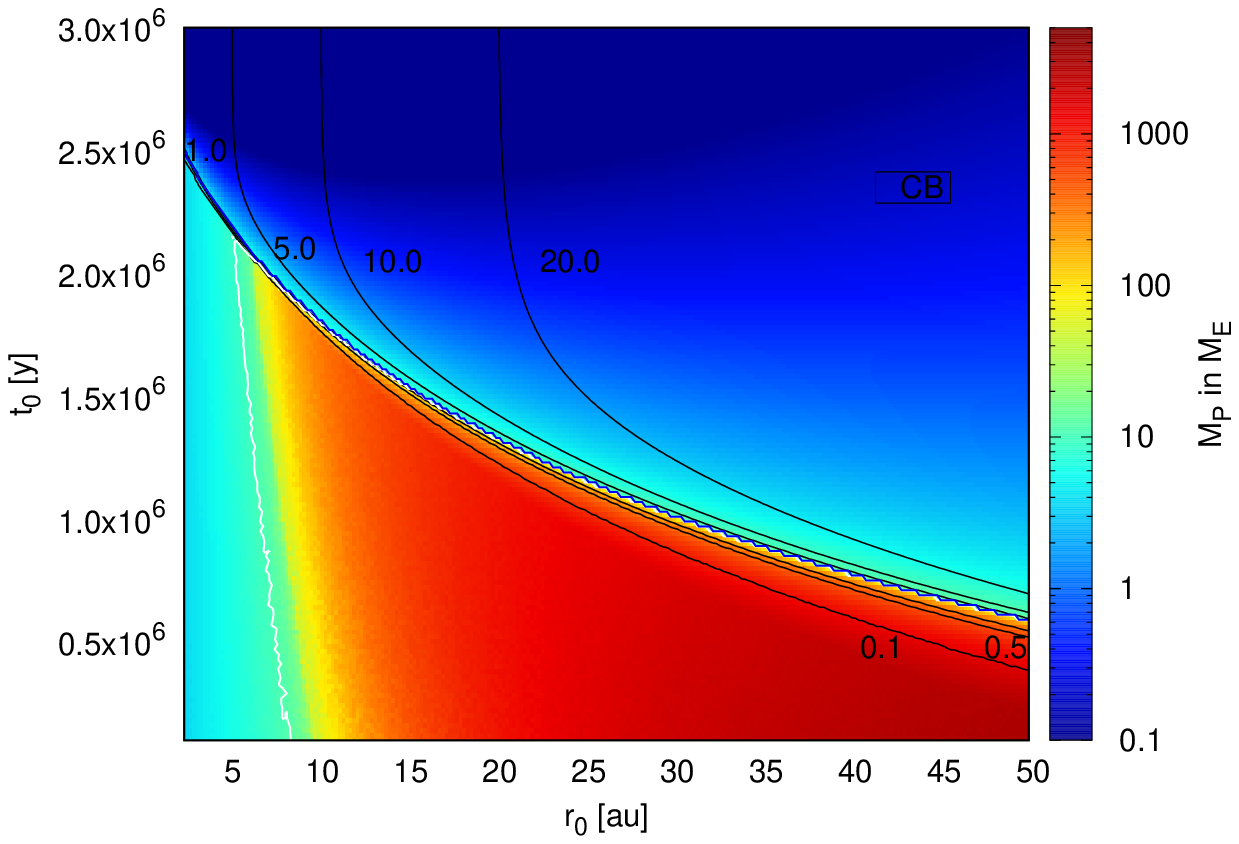} \quad
         \includegraphics[width=\columnwidth, height=2.45in]{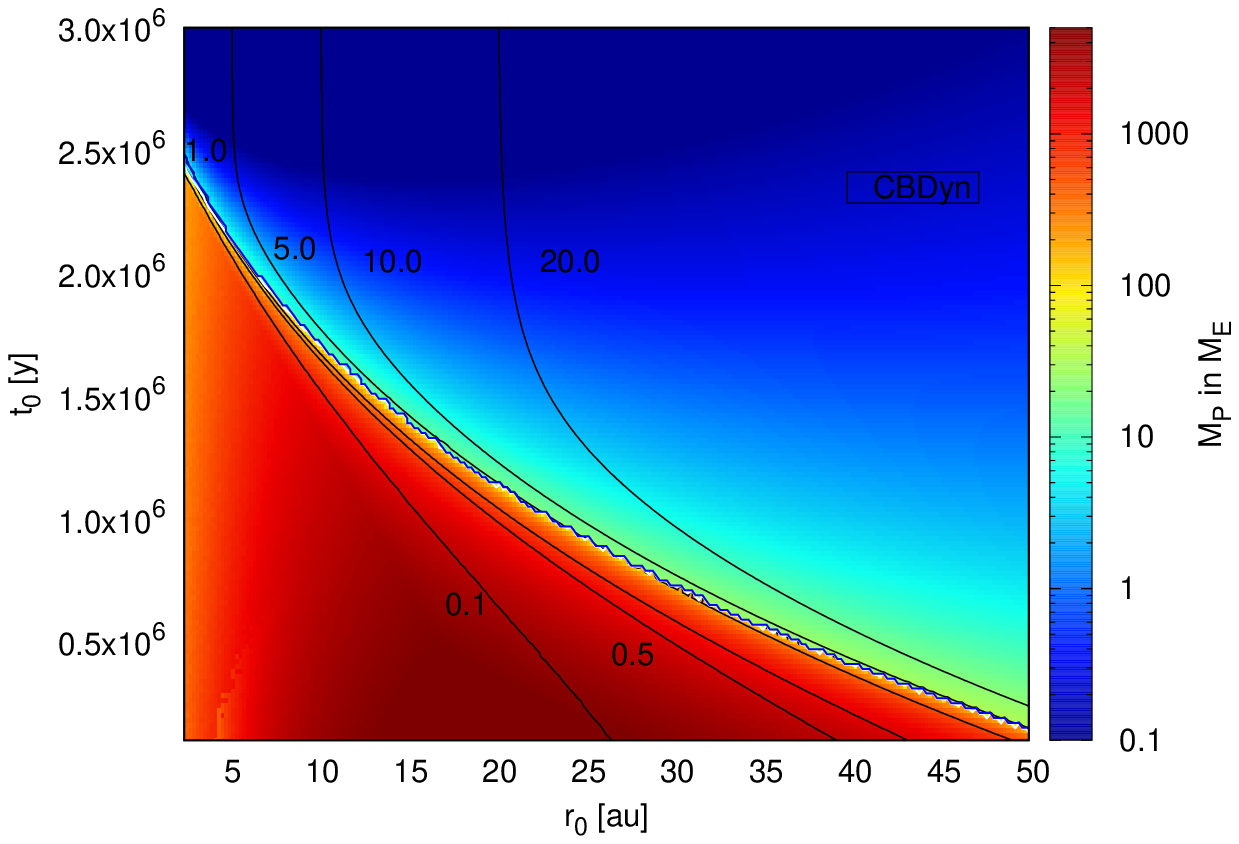}
\caption{Final masses of planets as a function of initial time ($t_{0}$) and and initial location ($r_{0}$) of their planetary embryos. Here the colours indicate final planetary masses. The black curves depict the final orbital positions, the blue line marks the pebble isolation mass, where planets below the line have reached pebble isolation mass. All planets inside the white line are in the runaway gas accretion regime with $M_{\rm core} < M_{\rm env}$. The first row plots feature NN (first column plot) and NNDyn (second column plot). The second row plots feature CB (first column plot) and CBDyn (second column plot).}
         \label{fig:2}
 \end{figure*}
consistency reasons, as we compared the results to the nominal models of \cite{Ndugu2018} that did not incorporate the migration prescription of \cite{Kanagawa2018}.
The contribution of the envelope opacity ($\kappa_{\rm env}$) to the planetary growth rate is reflected in the envelope contraction rates, where we use the gas envelope contraction rates derived from \cite{Ikoma2000} 
\begin{equation}  
\label{Ikoma2000}
{\dot{M}_{\rm gas, Ikoma}} =  \frac{M_{\rm p}}{\tau_{\rm KH}}.
\end{equation} Where, $\tau_{\rm KH}$ is the Kelvin-Helmholtz contraction rate and  scales:

\begin{equation}
{\tau_{\rm KH} }= 10^{3}\left(\frac{M_{\rm c}}{30 M_{\rm E}}\right)^{-2.5}\left(\frac{\kappa_{\rm env}}{0.05 \rm cm^{2}g^{-1}}\right){\rm year}.
\end{equation} Here $M_{\rm c}$ is the mass of the planet's core, in contrast with $M_{\rm p}$ which is the full planet mass (core + envelope). An alternative formulation of the gas-accretion rate was provided by \cite{Machida2010} as the minimum of:
\begin{equation}
 \dot{M}_{\rm gas,\rm low} = 0.83\Omega_{\rm k}\Sigma_{\rm g}H^{2}\left(\frac{r_{\rm H}}{H}\right)^{\frac{9}{2}}
 \label{Machida1}
\end{equation}
and 
\begin{equation}
 \dot{M}_{\rm gas,\rm high} = 0.14 \Omega_{\rm k}\Sigma_{\rm g} H^{2} \ .
 \label{Machida2}
\end{equation} The two different regimes originate from different planetary masses, one where the planetary Hill radius is smaller than the disc's scale height ($\dot{M}_{\rm gas,\rm low}$) and the other where the planetary Hill radius is larger than the disc's scale height ($\dot{M}_{\rm gas,\rm high}$). The \cite{Machida2010} rate is derived from shearing box simulations, where gap formation is not taken fully into account. However, once a gap is opened, obviously the planet cannot accrete more gas than the disk can supply. Throughout our simulations, we modeled the disc supply rate
\begin{equation}\label{disc}
\dot{M}_{\rm disc}=0.8\times3\pi\alpha H^{2}\Omega_{\rm K}\Sigma_{\rm g},
\end{equation} where $H$ and $\Sigma_{\rm g}$ are the disc scale height and gas surface density, respectively.  The pre-factor $0.8$ accounts for the fact that only 80~\% of the gas flux from the disc can be accreted by the giant planet, as found in hydrodynamical simulations \citep{Lubow2006}.  $\alpha$ sets the accretion flow in the disc, so it provides the gas accretion rate to the planet. Therefore, a change in $\alpha$ would imply a different accretion flow through the disc and thus sets a different limit on the planet accretion rate. In summary, our gas accretion rate, ${\dot{M}_{\rm gas}},$ onto the planet is taken as

\begin{equation}\label{gasaccrate}
 \dot{M}_{\rm gas} = \min{\left( \dot{M}_{\rm gas, Ikoma}, \dot{M}_{\rm gas, Machida},\dot{M}_{\rm disc}+f_{\rm HS}\dot{M}_{\rm HS}\right)},
\end{equation} where $ \dot{M}_{\rm HS}$ is the horseshoe depletion rate (explicitly introduced in section~\ref{Gapopening}). $f_{\rm HS}$ is a parameter which is set to 0 if the accretion of gas from the horseshoe region is not accounted for and 1 if it is.

We stop the simulation when the planet either (i) reaches 0.04~au, where we assume migrating planets are trapped at the inner disk edge \citep{Masset2006,Flock2019} or (ii) at the time of disc dispersion for the planets that do not reach the inner edge. All the models studied in this paper are described in Table~\ref{tab:ambiguous}.

Each of the blue curves in Figure~\ref{fig:1} traces a single evolution for a planet and differ for the choice of $\alpha$, and the inclusion or not of gas accretion from the horseshoe region and of the dynamical torque. Repeating the solid blue curve simulation in Figure~\ref{fig:1} for a range of planetary injection time and positions, we come up with a growth evolution map shown in the first column of the top panel of Figure~\ref{fig:2}. For this approach we vary the initial assumptions: starting time [0.1~Myr,~2.9~Myr] and starting position [0.2~au,~50~au]. We remind the reader that the initial time is a free parameter of the model, because it corresponds to the time when the planet seed achieves a mass corresponding to the transition between the Bondi and the Hill regime in the pebble accretion process at the planet location in the disc.

\section{Gap opening by Accretion} \label{Gapopening} 
Previous studies of planet-disc interactions only considered gap opening via gravitational perturbations \citep{Lin1986a,Crida2006,Crida2007}. Studies by \cite{Crida2017}, and subsequently by \cite{Bergez2020arXiv201000485B}, found that gas accretion by the planet in the runaway regime (third phase of the \cite{Pollack1996} scheme) can change the picture of planet formation significantly, in particular in terms of gap opening and accretion rate. They found using 2D isothermal hydrodynamical simulations that the accretion flow onto the planet comes from the neighborhood of the separatrix around the horseshoe region. This has two consequences that change the picture of the transition from Type~I to Type~II migration.
\begin{enumerate}
\item As the separatrix empties because of the accretion by the planet, the 
horseshoe region becomes partially isolated from the inner and outer discs. The flow from 
the inner and outer discs towards the separatrix is intercepted and accreted by 
the planet before it can reach the horseshoe region, if the accretion rates onto the planet are large. Hence, if the horseshoe 
region empties (because of the planetary accretion, see below), a gap is 
maintained, even if the planet is not massive enough to maintain a gap opened 
by its sole gravity according to the \cite{Crida2006} criterion.
\item As the separatrix empties, the horseshoe region diffuses viscously and 
tends to replenish the separatrix. This empties progressively the horseshoe region 
while its mass is transferred to the planet. In fact, the horseshoe material 
can be seen as a reservoir of mass that the planet can easily accrete, before 
being limited to what the disc can provide.
\end{enumerate}
These two effects combine to limit the transition phase between Type~I and 
Type~II migration, which is the most dangerous phase in terms of amplitude of 
inward migration. Not only can a planet reach the gap opening mass more 
rapidly, but it can actually open a gap before the classical gap opening mass, 
if the accretion rate is fast enough. Our model allows an inflow into an expanding horsehoe region by adding fresh material in the horseshoe mass reservoir. For a rapid expansion of the horseshoe region, like in 3D hydrodynamic simulations, new neighbouring gas might be included in the horseshoe region. Accreting this gas instantaneously onto the planet would not be physical. By adding it into the HS region, we allow for its accretion onto the planet on a synodic timescale. We explain the implementation of this finding in the following subsection.

\subsection{Implementation of the new gap opening and accretion model}\label{gapaccimplement}

Motivated by the recent isothermal simulation by \cite{Crida2017}, we assume that the planet accretes mass from the horseshoe region at the rate
\begin{equation}
 \dot{M}_{\rm HS}=M_{\rm HS}/(2T_{\rm HS}),
\end{equation} where $M_{\rm
  HS}$ is the mass of the horseshoe region, $T_{\rm HS}=3\pi\Omega
r_{\rm HS}/a$ is the synodic period at its border, $r_{\rm HS}=1.16 a
\sqrt{q/h}$ is its half-width, $q$ is the mass of the planet relative
to the mass of the star and $h$ is the aspect ratio of the disc at the
planet's location $a$.

At each time step $\delta t$ we compute the mass accretion rate $\dot{M}_{\rm HS}$ that could be provided by the horseshoe region. In the previous pebble-based planet formation model of \cite{Bitsch-etal-2015}, a ``depth of the gap'' parameter, $\cal{P}$  from \cite{Crida2007} was used.  $\cal{P}$ scales as 
 
\begin{equation}
\label{eq:gapopen}
 {\cal{P}} = \frac{3}{4} \frac{H}{r_{\rm H}} + \frac{50}{q \cal{R}},
\end{equation} where $r_{\rm H}$ is the Hill radius, $q=M_{\rm P} / M_\star$, and $\mathcal{R}$ the Reynolds number given by $\mathcal{R} = r_{\rm P}^2 \Omega_{\rm k} / \nu$. Here $\nu$ is the disc viscosity. ${\cal{P}}$ is an input for the parameter, $f({\cal{P}})$ representing the gravitational opening of the gap. $f({\cal{P}})$ is given by
\begin{equation}
          f({\cal{P}}) = \left\{
     \begin{array}{lr}
       \frac{\left({\cal{P}}-0.541\right)}{4}      &   : {\cal{P}} \leq 2.4646\\
        1-\exp\left(-\frac{{\cal{P}}^{0.75}}{3}\right) & : {\rm Otherwise}
     \end{array}.
   \right.
\end{equation}  Through $f({\cal{P}})$, the density of gas in the gap is computed as\\ $\Sigma_{\rm gap}={f(\cal{P})}\hat{\Sigma}_{\rm HS}$. Here $\hat{\Sigma}_{\rm HS}$ is the density of gas that the horseshoe region would have if gap opening did not occur. It may be different from the local unperturbed gas density in the disk, $\Sigma_{\rm disc} (a_{\rm p})$ because a migrating planet carries its horseshoe material with it. We return to this below. Following the same philosophy, we introduce an additional parameter $f_{\rm A}$, initially equal to 1, which is computed at every timestep. $f_{\rm A}$ scales: 
\begin{equation} \label{eq:gapacc}
f_{\rm A} = 1-\frac{{\dot{M}_{\rm gas}}\delta{t}}{f({\cal{P}})\hat{M}_{\rm HS}}.
\end{equation}
$\hat{M}_{\rm HS}$ is the mass inside the horseshoe region in the absence of gas accretion onto the planet and in absence of gravitational gap opening. $\hat{M}_{\rm HS}$ is given by
\begin{equation}
 \hat{M}_{\rm HS} =2\pi a r_{\rm HS} \hat{\Sigma}_{\rm HS}. 
\end{equation}  The full depth of the gap is therefore
 \begin{equation}\label{gapeff}
 f_{\rm gap}={f(\cal{P})} f_{\rm A}.
\end{equation} As in NN, when a partial gap is opened ($ 0.53<f_{\rm gap}<1$) the Type~I migration rate is reduced by the factor $f_{\rm gap}$. When $f_{\rm gap}<0.1$ the planet migrates in Type~II mode at the viscous accretion rate of the disc and, when $ 0.1<f_{\rm gap}<0.53$ the migration rate is computed as a linear interpolation in $f_{\rm gap}$ between the rates corresponding to $f_{\rm gap}=0.53$ and $f_{\rm gap}=0.1$. $f_{\rm gap}$ is the gap depth parameter. We note here that the goal of the linear interpolation is to provide a smoother transition between the classical type~I and type~II migration fashions \citep[e.g.][]{Dittkrist2014}. This has important consequences for planet migration, right from the time when planets start to carve small gaps in the disc. We remind the reader that with the effective gap parameter, $f_{\rm gap}$, the planet transitions into slower type~II migration faster, since  $ f_{\rm gap} < 0.1 $ occurs earlier compared to the classical case $f({\cal{P}}) < 0.1$.

The new formulae above (Equation~\ref{eq:gapacc}) requires us to monitor the mass of the horseshoe region 
$M_{\rm HS}$ as a function of time. By definition of the $f(\cal{P})$ and $f_{\rm A}$ 
factors, the mass of the horseshoe region scales as
\begin{equation}\label{HSmass}
M_{\rm HS}={f(\cal{P})}f_{\rm A}\hat{M}_{\rm HS}.
\end{equation}
The quantity $\hat{M}_{\rm HS}$ evolves over time because the width of the horseshoe region $r_{\rm HS}$ changes with the planet mass $q$ and location $a$. For simplicity, we assume 
that the vortensity in the horseshoe region is conserved (strictly speaking this is true only in the limit of vanishing viscosity), so that if the location of the planet changes
from $a$ to $a'$, the horseshoe gas density changes from 
\begin{equation}
\hat{\Sigma}_{\rm HS}=\hat{M}_{\rm HS}/(4\pi a r_{\rm HS})
\end{equation} to 
\begin{equation}
\hat{\Sigma}'_{\rm HS}=\hat{\Sigma}_{\rm HS}(a/a')^{3/2}.
\end{equation}
\begin{figure}
 \centering
 \includegraphics[width=\columnwidth]{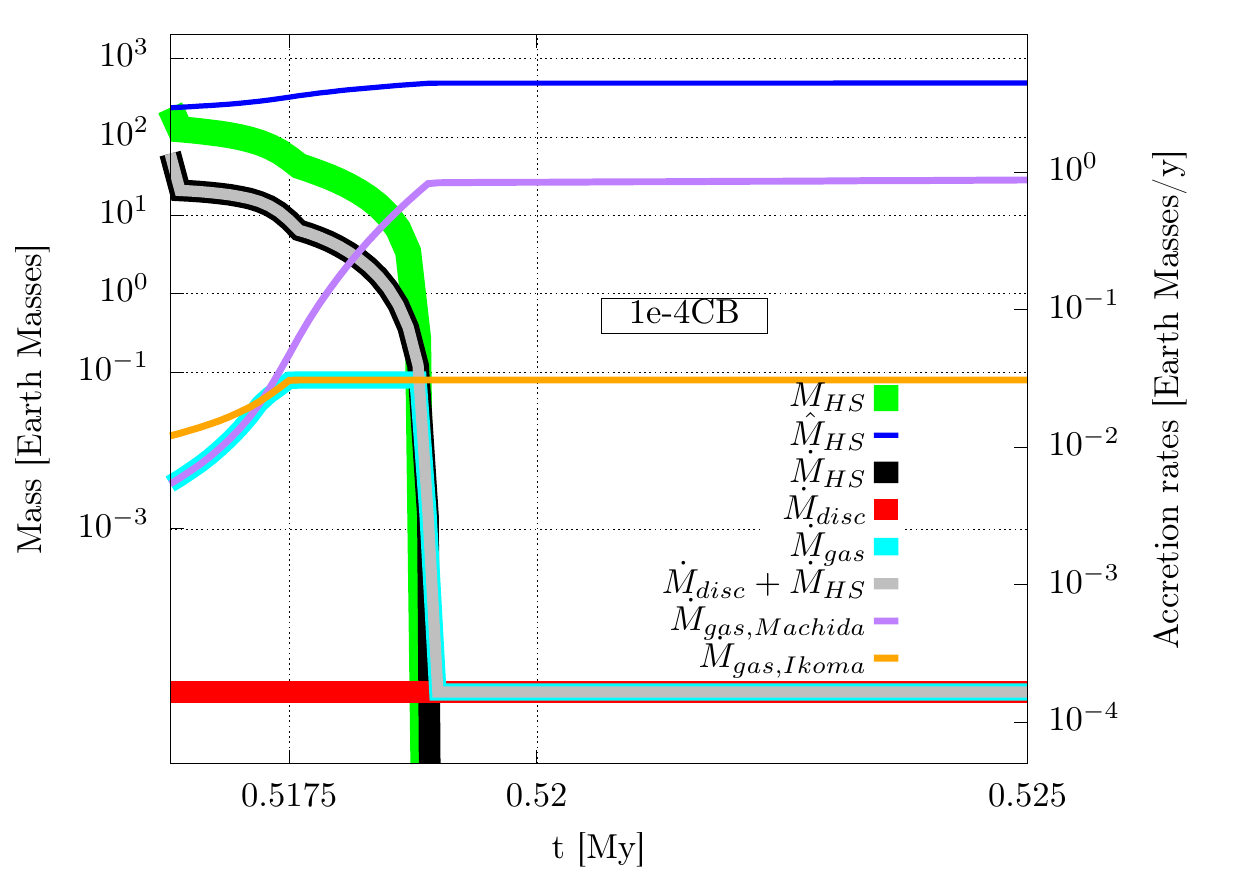}
 \caption{Time evolution of horseshoe gas mass (green curve) and the relevant gas accretion rates (from the horseshoe region - black; from the disk's viscous accretion - red; the sum of the two - gray; the effective gas accretion rate onto the planet - light blue). For simplicity we assumed that the planet has a fixed position of $a=10$~au and the disc does not evolve over time. The plots feature $\alpha-$viscosity value of 0.0001 and core mass of 16.56 Earth masses. The dark blue curve depicts the mass inside the horseshoe region without taking any depletion by accretion into account.}
  \label{fig:1massconserv}
\end{figure}
Thus, when $r_{\rm HS}$ changes to $r'_{\rm HS}$, we compute the quantity

\begin{figure}
 \centering
 \includegraphics[width=\columnwidth]{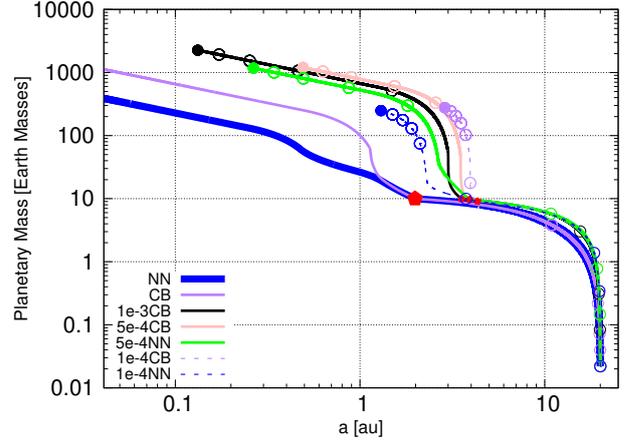}
 \caption{Growth track showing the impact of $\alpha$-viscosity on the mass-distance distribution of forming planets. With the exception of planetary embryos implanted at 20~au,  the plots are obtained with the same initial starting time (0.1~Myr) and disc dispersion time (3~Myr) as in Figure~\ref{fig:1}. The marks on the curves have same meaning as in Figure~\ref{fig:1}. The sub models are defined in Table~\ref{tab:ambiguous}}
  \label{fig:1b}
\end{figure} \begin{equation} \hat{M}'_{\rm HS}=4\pi a'r'_{\rm HS} \hat{\Sigma}'_{\rm HS}. 
\end{equation} If $\hat{M}'_{\rm HS}<\hat{M}_{\rm HS}$, we refill the horseshoe region at the disc's viscous spreading rate and recompute $\hat{M}'_{\rm HS}$ as
\begin{equation}
 \hat{M}'_{\rm HS} =  \hat{M}_{\rm HS} + (\dot{M}_{\rm disc}-\dot{M}_{\rm gas})\delta t, 
\end{equation} where $\dot{M}_{\rm disc}$ and $\dot{M}_{\rm gas}$ are defined in Equation~\ref{disc} and Equation~\ref{gasaccrate}, respectively. If the opposite is true, it means that the horseshoe region has expanded and must have captured new gas from the disc, 
with a density $\Sigma_{\rm disc}$. Thus, we compute the new value of $\hat{M}_{\rm HS}$ as:
\begin{equation}\label{eq:newMH}
\hat{M}'_{\rm HS}=\hat{M}_{\rm HS} + \left(4\pi a' r'_{\rm HS} -{{\hat{M}_{\rm 
HS}}\over{\hat{\Sigma}'_{\rm HS}}}\right) {\Sigma_{\rm disc}}\ .
\end{equation} Once $\hat{M}'_{\rm HS}$ is computed, the new value of $\hat{\Sigma}'_{\rm HS}$ is recomputed as $\hat{\Sigma}'_{\rm HS}=\hat{M}'_{\rm HS}/(4\pi a' r'_{\rm HS})$. This procedure is then repeated at every time step. This procedure automatically captures the gas surface density decay during the disc's evolution, because 
$\Sigma_{\rm disc}$ is evaluated at each time step. We point out here that Equation~\ref{eq:newMH} assumes that outside of $r_{\rm HS}$, the surface density of the disc is unpertubed. This is equivalent to assuming that the  gap profile is a heavyside function of width $r_{\rm HS}$. We acknowledge that our approximation is crude for very massive planets, but should hold for intermediate mass planets and so for planets growing by gas accretion and migrating by Type~I and dynamical torques. In addition, the horseshoe refilling rate scales with alpha-viscosity. For example, at high $\alpha$ values, the horseshoe region would get refilled by the disc more efficiently because the disc supply rate is larger.
\begin{figure*}
        \centering
         \includegraphics[width=\columnwidth, 
height=2.45in]{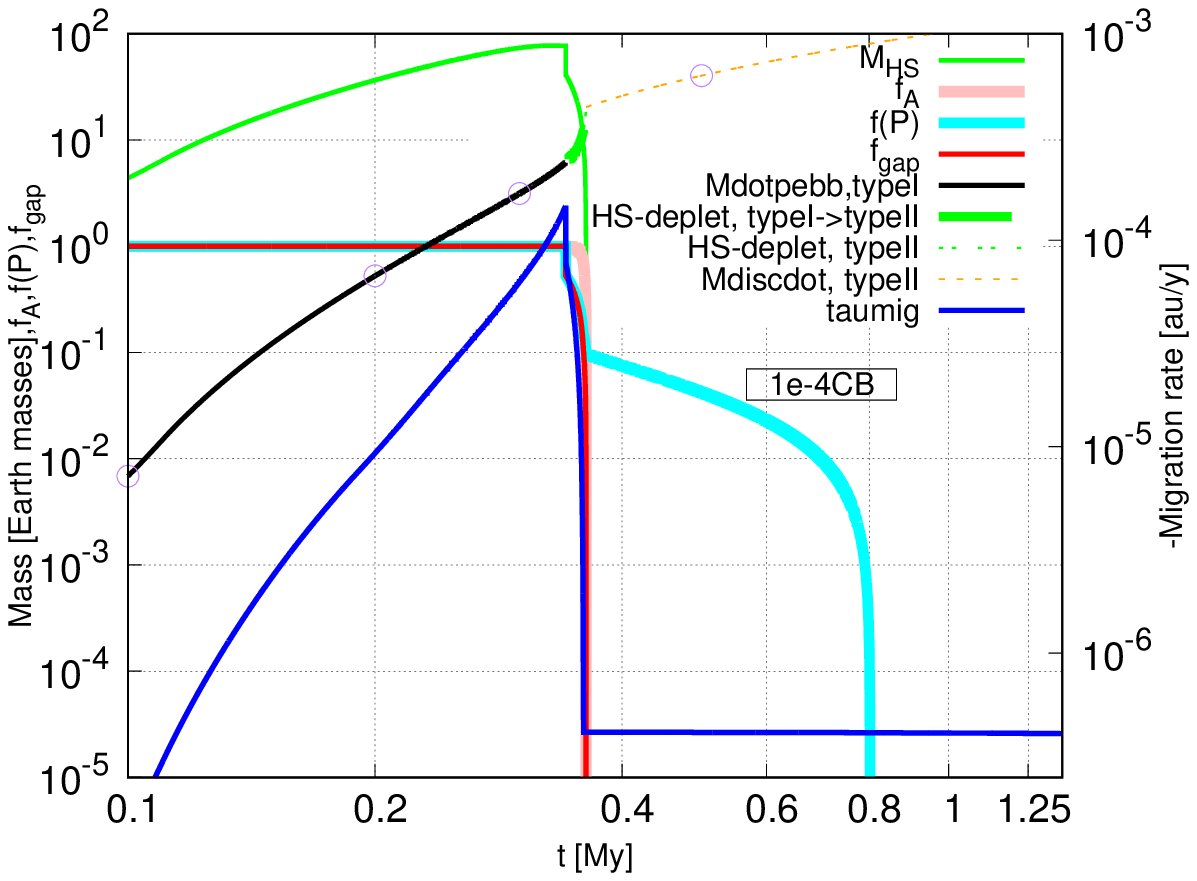}\quad
        \includegraphics[width=\columnwidth, 
height=2.45in]{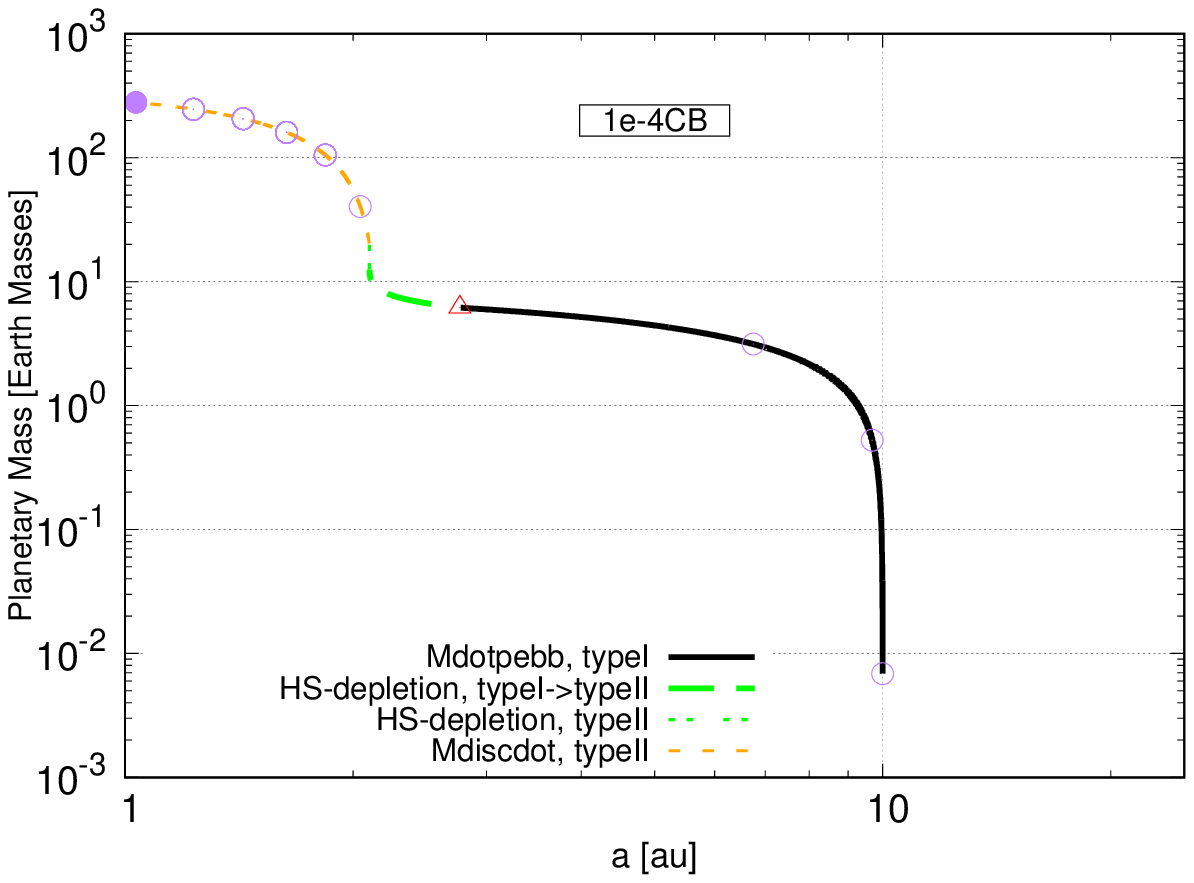} \quad
  \caption{The left panel shows the time dependency of the horseshoe gas mass (green curve) from Equation~\ref{HSmass}, gap depth parameter due to horseshoe gas accretion (pink curve) and due to gravitational pushing (light blue curve), effective gap depth parameter (red curve) from Equation~\ref{gapeff} and the inward migration rates (deep blue curve) for the 1e-4CB track displayed in Figure~\ref{fig:1}. The left panel plot also shows the $m-t$ relationship for 1e-4CB track (pebble accretion (black curve), horseshoe gas depletion (dashed green curve) and when the disc's accretion rate limits the accretion onto the planet (orange dashed curves). The right panel shows the main sequence of events occurring in the $m-a$ trajectory of the 1e-4CB displayed in Figure~\ref{fig:1}. The planet accretes pebbles and concurrently undergoes migration in the type~I regime (black curve). After the planet reached the pebble isolation mass (red triangle), the planet starts accreting gas from the horseshoe region, causing its depletion (green curve). The depletion regime of horseshoe gas corresponds closely to the transition to type~II migration. The core then undergoes runaway gas accretion in the regime marked by the dashed orange curve with the migration rates in this regime corresponding to the type~II migration.}
         \label{fig:1massmig}
 \end{figure*} Figure~\ref{fig:1massconserv} shows that in our approach, the horseshoe gas mass decay is due to the planetary gas accretion and gravitational pushing of the protoplanet. The final planetary gas mass by the time of the horseshoe depletion corresponds to the initial gas mass in the horseshoe region. Thus our simple approach of horseshoe gas accretion is mass conservative.

\subsection{Consequences on planet formation} \label{consequences}
Using the previous outlined modifications, we plot the growth track of a 
planet starting at $a_{\rm 0}=10$~au and $t_{\rm 0}=0.1$~Myr as a purple line in Figure~\ref{fig:1}. For the nominal  $\alpha$ viscosity reported in this work (0.005), the discussed new prescription of gas accretion appears irrelevant, especially in slowing the overall migrating distance to the star. We attribute this result to the faster type~II migration rates due to the high $\alpha$-viscosity. In addition, the high $\alpha$-viscosity means high pebble scale height, therefore the pebbles are not in the efficient accretion zones of the planetary embryo. Therefore migration wins over growth, with the planet starting to carve a gap when the cores is approaching 1~au from the star and migrates even further inwards before transitioning fully into slower type~II regime. We test the importance of core location in the disc by implanting cores at 20~au in Figure~\ref{fig:1b}. As expected the core now starts carving gap just outside of 2~au but the planets still migrate all the way to the disc's inner edge. We therefore conclude that, within the framework of the simple power law disc model we used, the difference between NN and CB in slowing inward migration is not very significant for higher $\alpha$ viscosity.

Assuming a lower $\alpha$-viscosity (0.0001) in the disc, we clearly see differences appearing between NN and CB model in Figure~\ref{fig:1}. Both 1e-4NN and the 1e-4CB simulations now accrete pebbles in the efficient 2D regime. This allows the early formation of a gap, before the planet is driven to the inner edge of the disc. The 1e-4CB curve (short dashed purple) starts to diverge relative to the 1e-4NN model (short dashed blue curve), when the planet starts its runaway gas accretion. From this point on, the growth is faster in the 1e-4CB case, because gas accretion is not limited by the disc accretion as long as the horseshoe region is not empty. As soon as the horseshoe region empties, the growth slows down and becomes limited by what the disc can provide. At the same time the planet transitions into Type~II migration. Therefore, the intermediate phase between Type~I and Type~II migration is very short. At the end of the disc lifetime the planet following the 1e-4CB track ends just outside 1~au and with a mass of around 300 Earth masses. For these particular growth tracks (Figure~\ref{fig:1}), an important difference between 1e-4NN and 1e-4CB is noted in the final planetary orbital positions, but only a small difference is seen in terms of the final planetary mass. Comparing Figure~\ref{fig:1} and Figure~\ref{fig:1b}, we note that the difference between NN and CB prescriptions scales with: (i) $\alpha$ viscosity and (ii) injection positions of planetary embryos in the discs. The effect of the CB prescription is clearly highlighted for cores implanted in the outer disc. In this prescription the gas surface density of the protoplanetary disc plays a crucial role in the final mass of the planet, as it determines how much material the 
\begin{figure*}
        \centering
         \includegraphics[width=\columnwidth, 
height=2.45in]{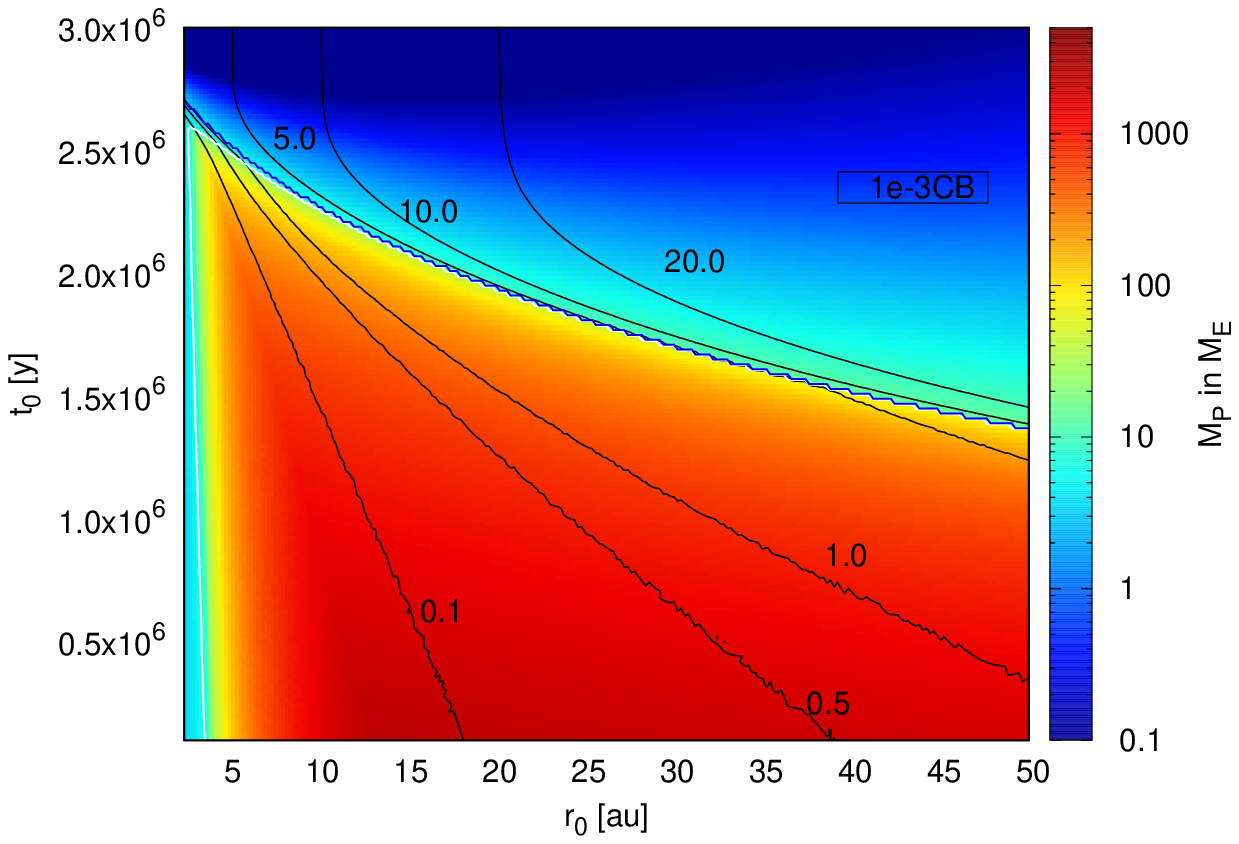}\quad
        \includegraphics[width=\columnwidth, 
height=2.45in]{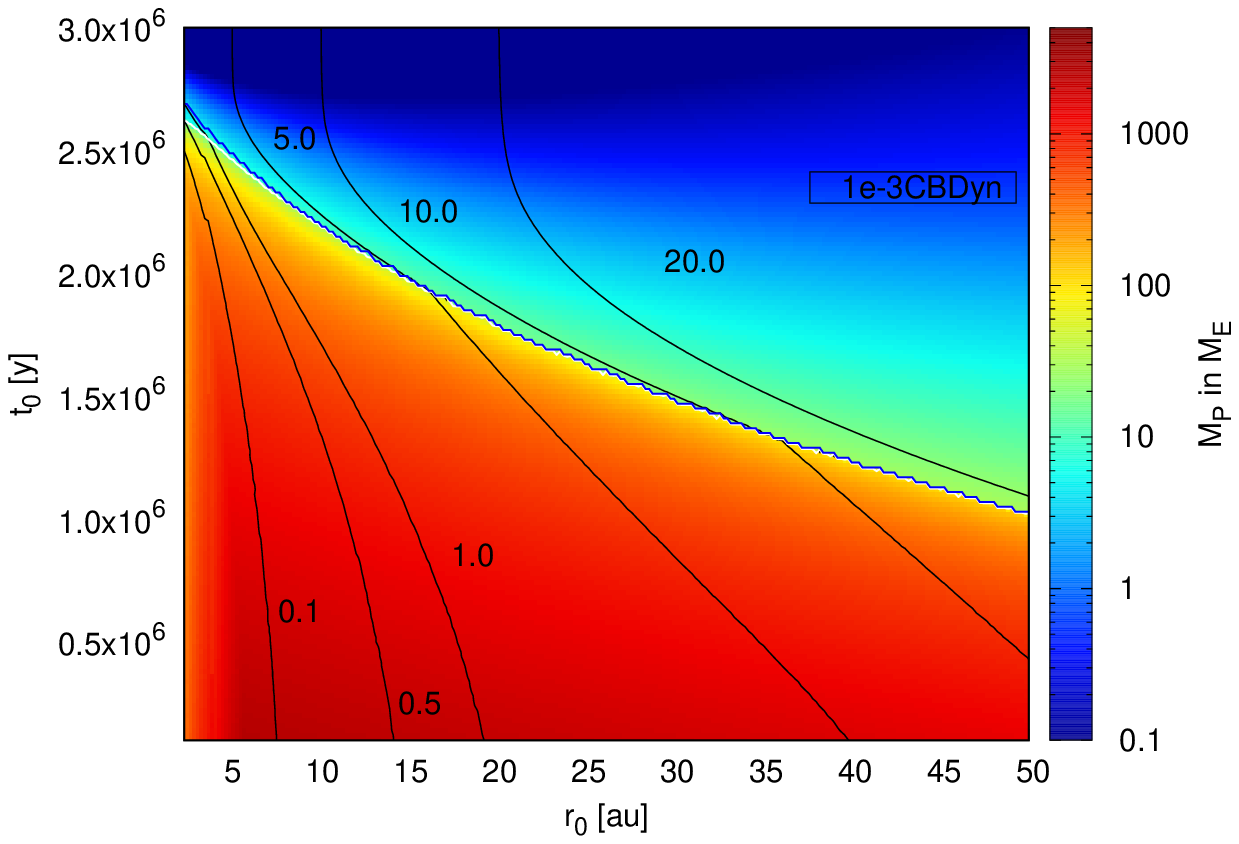} \quad
         \includegraphics[width=\columnwidth, height=2.45in]{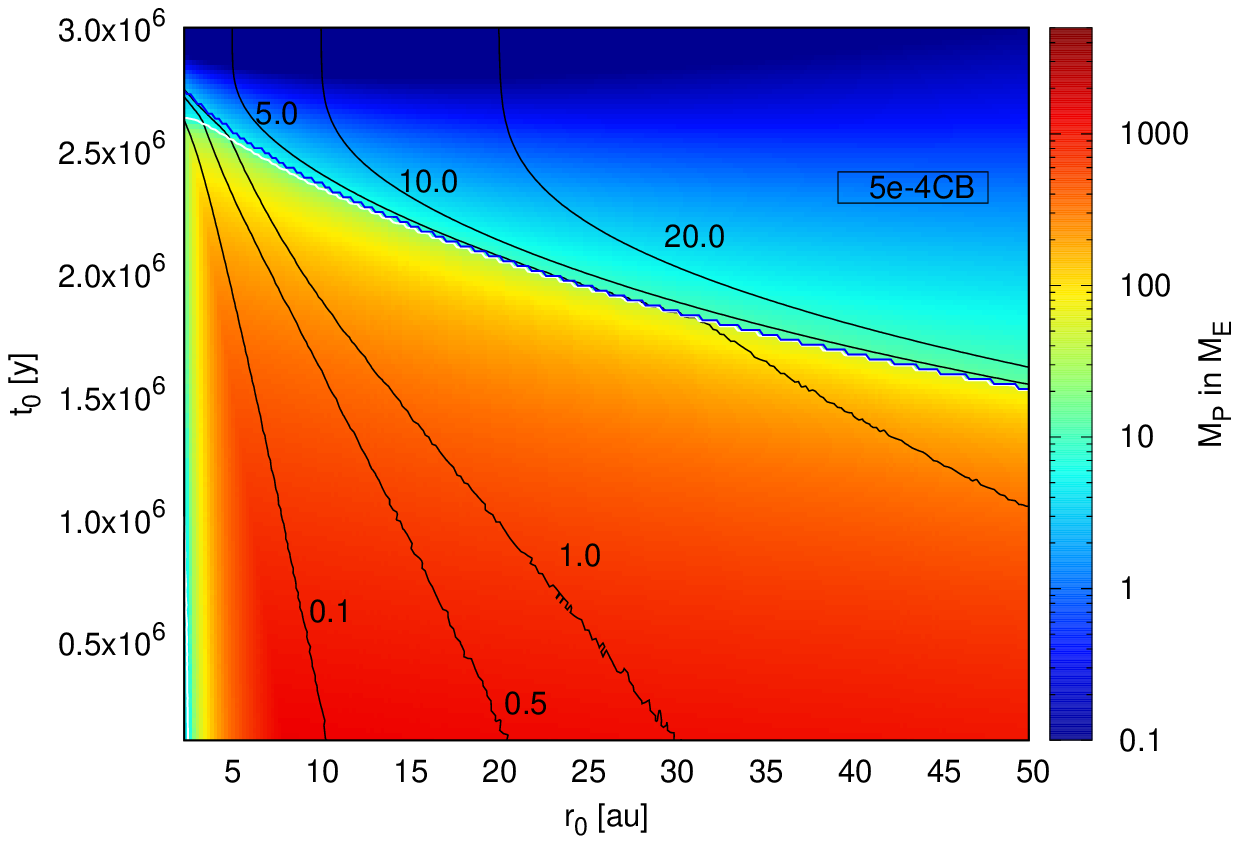} \quad
         \includegraphics[width=\columnwidth, height=2.45in]{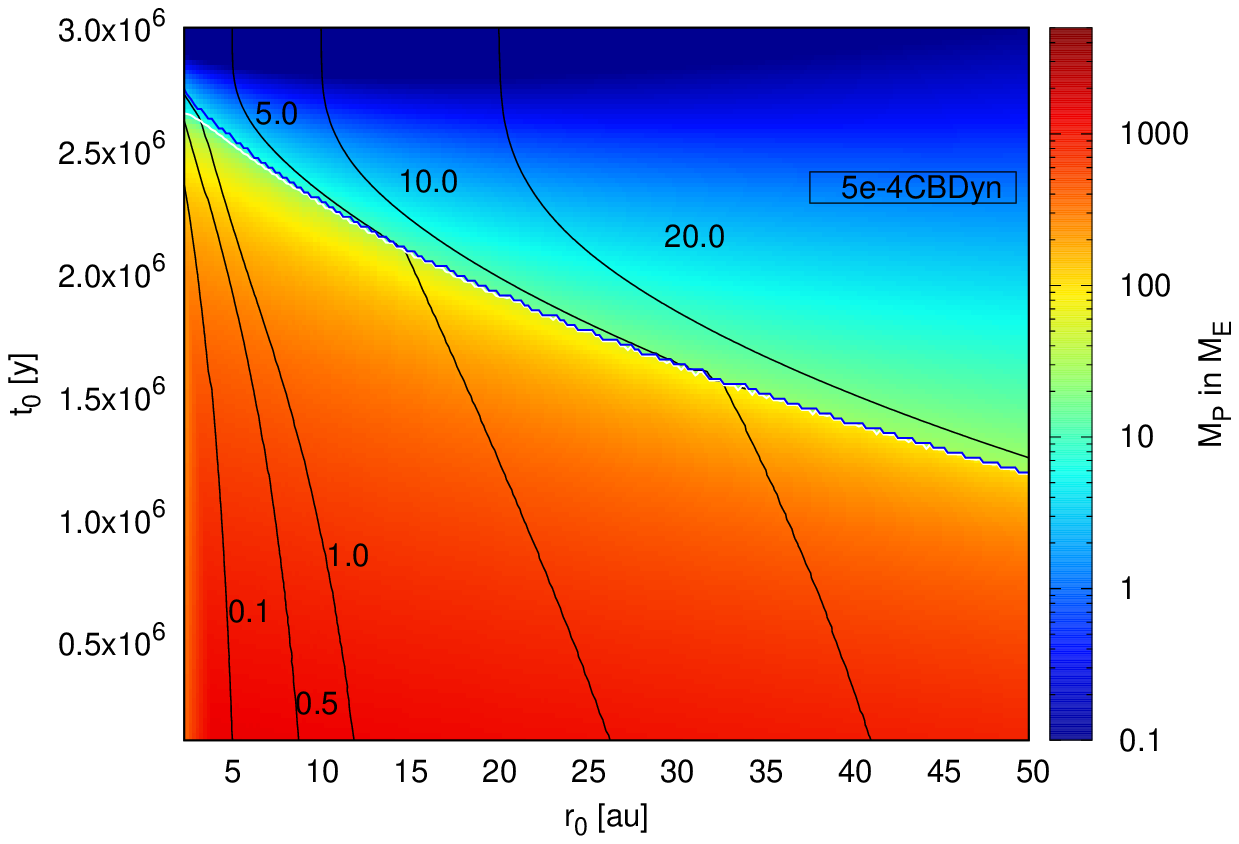}
         \includegraphics[width=\columnwidth, height=2.45in]{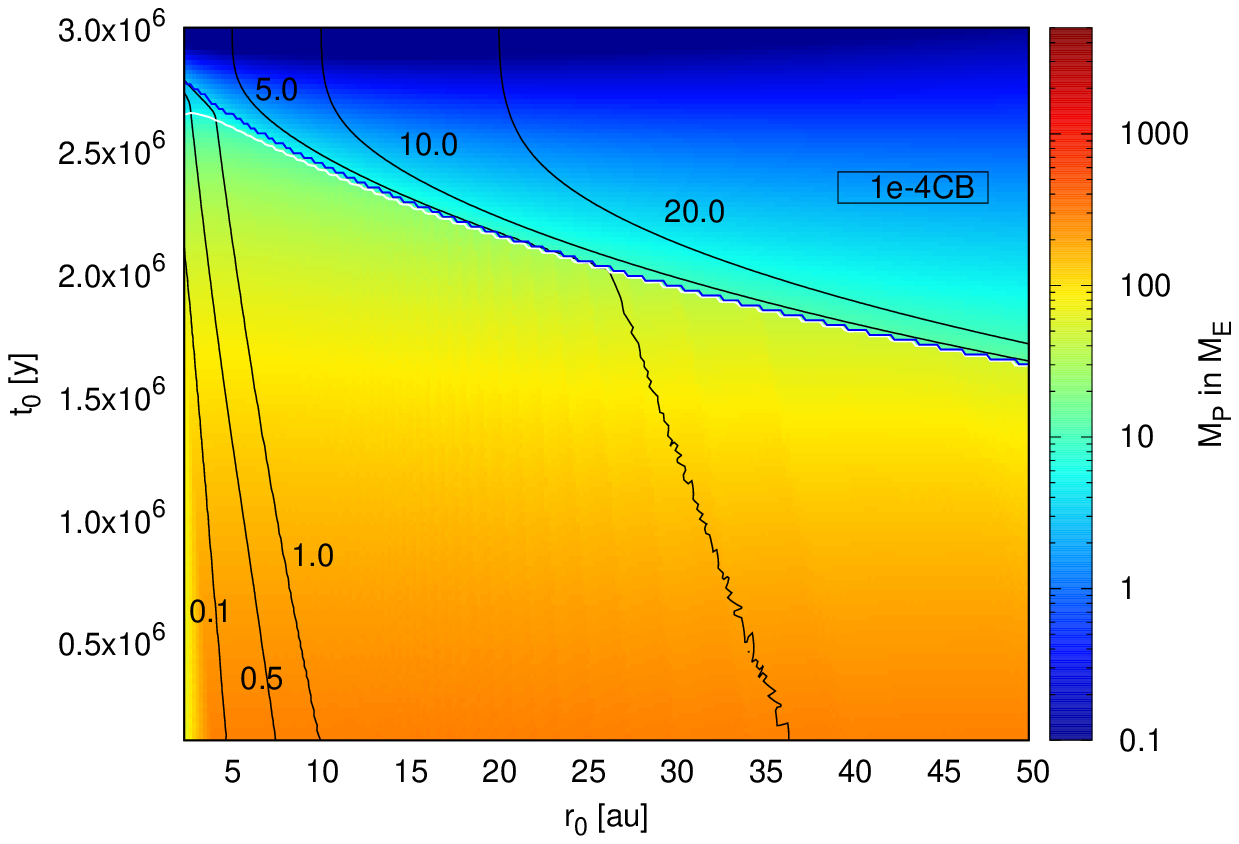} \quad
         \includegraphics[width=\columnwidth, height=2.45in]{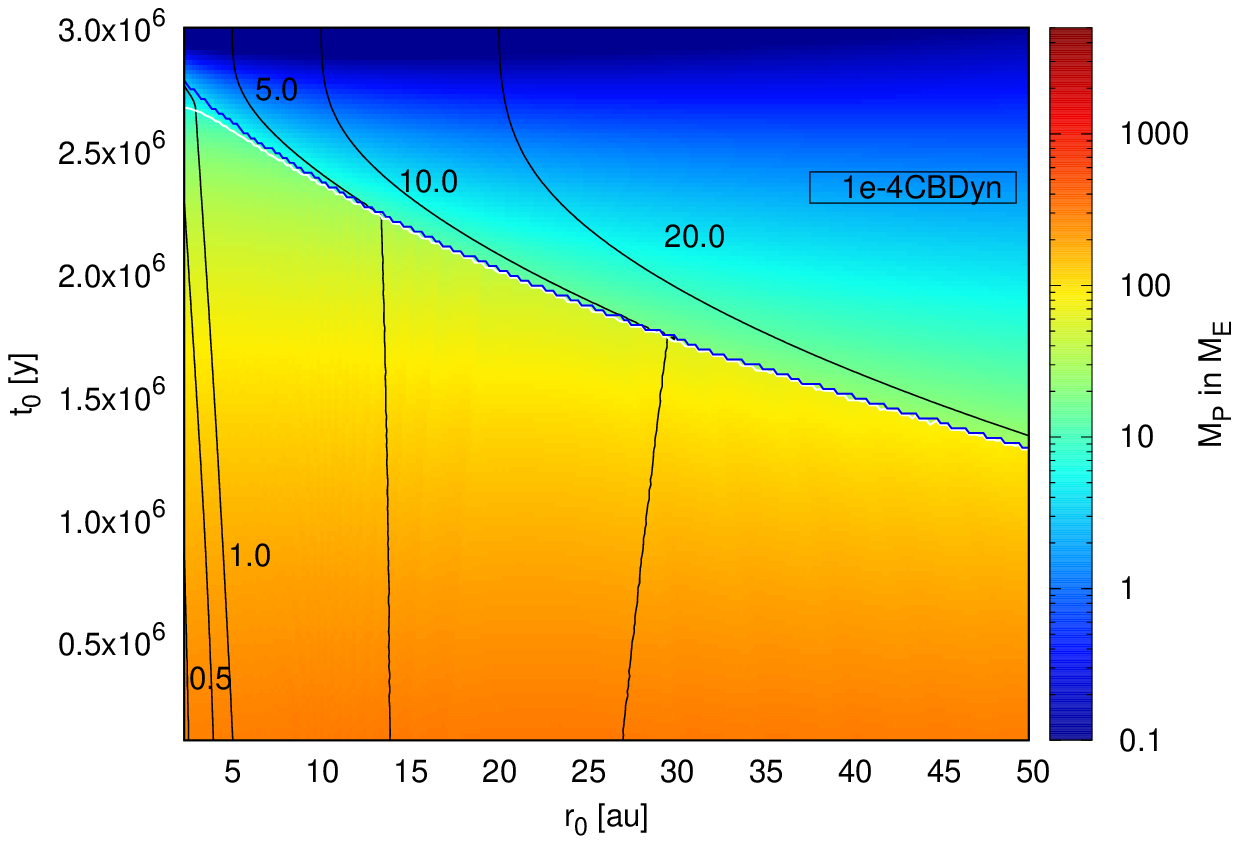}
  \caption{Growth map showing the impact of the disc's $\alpha$-viscosity parameter on planets growing with the CB prescription. The top, middle and bottom plots feature the 1e-3CB, 5e-4CB and 1e-4CB models. All the lines have the same meaning as in Figure~\ref{fig:2}, but we now feature different $\alpha$ values (top to bottom pannels) and include the effect of the dynamical torques (left pannel plots).}
         \label{fig:4}

\end{figure*} planet can accrete from its horseshoe region\footnote{The amount of materials in the core's horseshoe region depends on the accretional viscosity and the core's location in the disc.}. This is a clear difference with respect to the NN model, where the final planetary mass depends only on the mass accretion rate through the disc and not on the total gas surface density. Planets that formed earlier, when the disc is more 
massive, can grow larger than planets formed later. In addition, if planets are formed late, the differences between CB and NN should be very small due to the limited material available in the horseshoe region (see the global maps (Figures~\ref{fig:2},~\ref{fig:4}~\&~\ref{fig:5}) for more illustration). The corresponding global map for NN and CB models are shown in the first column of the top and the bottom panel 
of Figure~\ref{fig:2}, respectively. We see that there are many more super Jupiters in hot/warm orbits with respect to the nominal case (first column of the top panel of Figure~\ref{fig:2}), because the more rapid transition to Type~II migration reduces the overall migration of the planet towards the star. The planets, moreover, are a bit more massive than in the nominal case.

We show in Figure~\ref{fig:1massmig} the overall sequence of major events in the track 1e-4CB (short-dashed purple curve) of Figure~\ref{fig:1}. It is clear from Figure~\ref{fig:1massmig} that the planetary migration rate occurs in three regimes (marked by different colors). The type~I migration rate scales linearly with planetary mass until it opens a gap. The planet then gradually transitions into the constant slower type~II migration rate. We remind the reader that the onset of our horseshoe gas mass depletion corresponds to the intermediate type~I/type~II regime. In addition, our effective gap parameter profile (black curve) follows the horseshoe gas mass profile (green curve) and is only marginally determined by the gravitational gap opening, highlighting the role of gas accretion for gap opening.

We stress that the formula of \cite{Crida2007} for the gap depth would predict the opening of a gap with 50\% depth at $\sim 5$~Earth masses, determining the end of pure Type~I migration. However, such a gap 
would cause the blockage of the flux of pebbles at the outer edge of the gap. 
In other words, opening a gap at 5~Earth masses is inconsistent with estimate of 
the pebble isolation mass at $\sim 20$~Earth masses at disc location with $\frac{H}{r} 
= 0.05$ \citep{Bitsch2018}. We think that the \cite{Crida2007} formula, 
calibrated for giant planets in viscous discs, is not quantitatively reliable 
for small planets in disc of vanishing viscosity, as stated in 
\cite{Bitsch2018}. Instead, the pebble isolation mass criterion has been 
investigated in the proper regime by recent 
hydrodynamical simulations \citep{Lambrechts2014,Bitsch2018,Ataiee2018}. Thus, 
we assume that Type~I migration operates until the pebble isolation mass. 

In the transition period between Type~I and Type~II migration, the planet implanted at 10~au and at 0.1~Myr in a disc with $\alpha=0.0001$ using the CB accretion prescription (short dashed-purple line in Figure~\ref{fig:1}) achieves 90~\% orbital decay. Applying the dynamic co-rotation torque, we obtain the tracks depicted by the long dashed curves in Figure~\ref{fig:1}. Following the  purple long dashed curve (1e-4CBDyn track) in Figure~\ref{fig:1}, Type~I migration now brings the planet to $\sim 3$~au. 
 \begin{figure*}
        \centering
         \includegraphics[width=\columnwidth, 
height=2.45in]{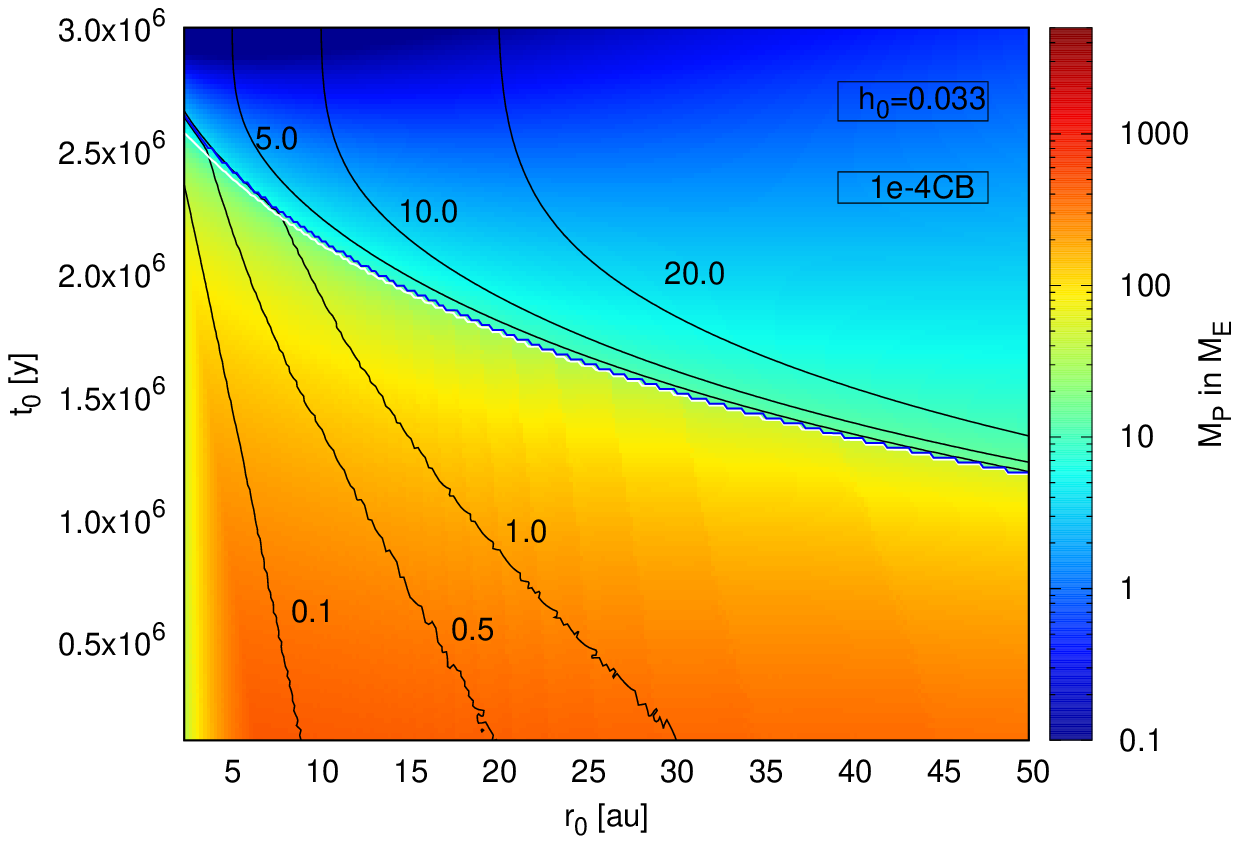}\quad
        \includegraphics[width=\columnwidth, 
height=2.45in]{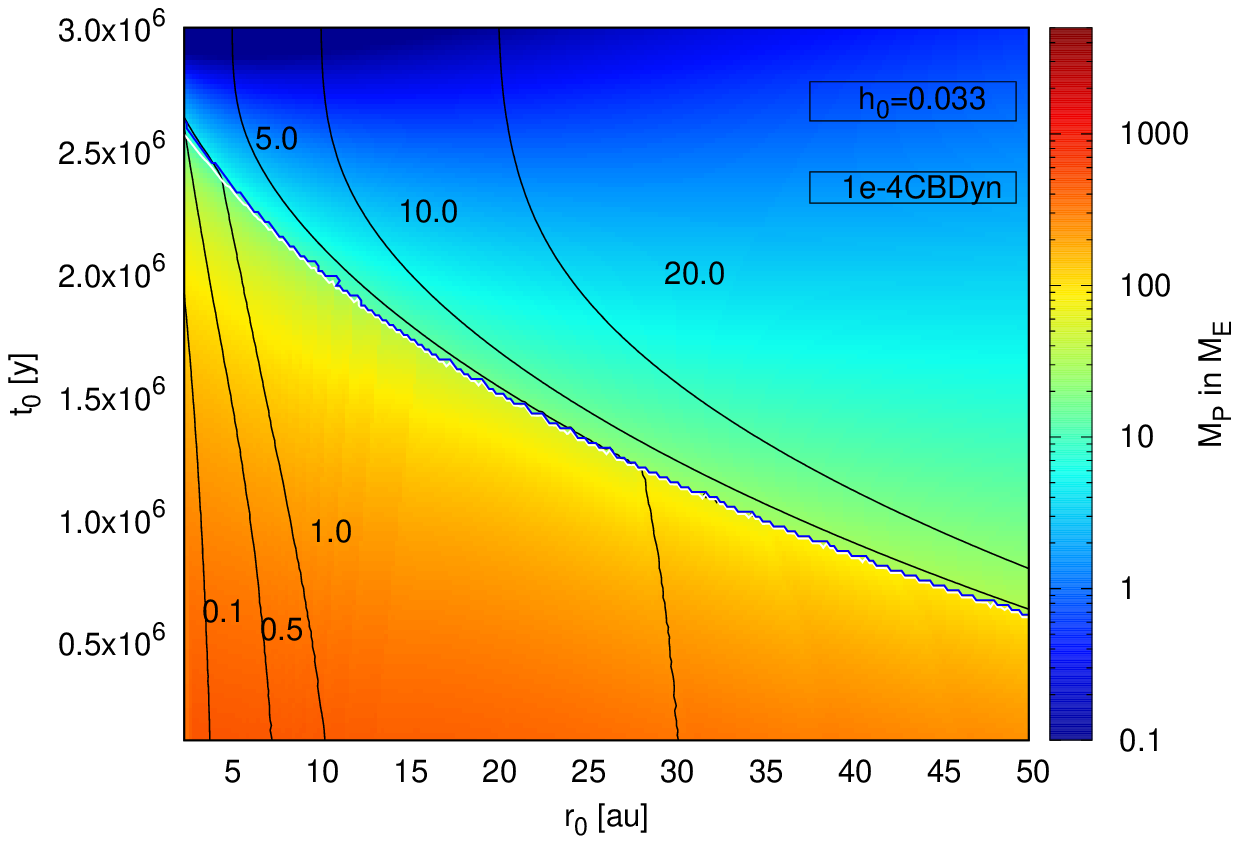} \quad
  \caption{ Growth map of planets growing in a disc with a higher aspect ratio ($h_0=0.033$). We show the results of the 1e-4CB (left) and 1e-4CBDyn (right) models. All lines have the same meaning as in Figure~\ref{fig:4}.}
         \label{fig:5}
 \end{figure*}
Runaway gas accretion therefore starts just before 0.4~Myr, when the planet is still beyond 4~au, bringing the planet into the Type~II migration regime. Then, as expected, due to the low viscosity of the disc, the migration is slow. In the subsequent 2.6~Myr the planet moves radially by $\sim 1$~au, finding itself at $\sim 300$~Earth masses just inside of 3~au at the end of the disc's lifetime.  This result confirms the relevance of the dynamical torque in planet formation.

\section{When are the CB accretion and the dynamical torques relevant in planet formation?} \label{when}
The previous section showed that, in the CB prescription and accounting for the dynamical corotation torque, the viscosity of the disc and the location and time of injection of the planetary seed in the disc play a major role. In this section, we show when and where the CB prescription and the dynamical torque become important in planet formation using a global evolution map of planet formation like those of Figure~\ref{fig:2}.
We start by exploring the impact of $\alpha$ viscosity in the disc on the global planet formation map. The $\alpha$ viscosity in our model controls the migration and accretion speed, for example at a higher $\alpha$ viscosity, inward migration is fast and the gas accretion rate of a planet exceeding the pebble isolation mass is high. In addition a lower viscosity reduces the pebble scale height, allowing faster growth during the pebble accretion stage.

It is clear from the simulations with $\alpha=0.0001$ that the lower migration speeds and faster pebble-accretion rates allow the formation of gas giants also in the inner parts of the protoplanetary disc (see the first column plots in Figure~\ref{fig:4}). On the contrary, for higher $\alpha$ viscosity, the  gas giant planets have to start forming in the disc outskirts due to the fast inward migration speed that outperforms growth rates (see the first column plots in Figure~\ref{fig:2}).

The effects of the CB prescription become enhanced at low viscosities. This results in earlier gap opening, less inward migration and slightly larger gas giants (e.g. Figure~\ref{fig:1} and Figure~\ref{fig:1b}). However, in the case of $\alpha=10^{-4}$, the gas accretion rate in itself is very low, resulting in planets that barely reach super Jupiter mass, indicating that a higher viscosity might be needed to explain very massive giant planets. In addition the effects of the CB prescription depends on the injection time of the planetary embryo. In particular, an early injection time of the planetary embryo results in a stronger effect of the CB accretion compared to later planetary embryo injection times. This is caused due to the fact that the mass of gas in the planet's horseshoe region is larger at earlier times.

We show in Figure~\ref{fig:4} (second column plots) the influence of the dynamical torque on the final masses and positions of planets formed in our model. Quite clearly, the dynamical torque becomes more efficient in slowing-down migration if the disc's viscosity is smaller \citep{Paardekooper2014}. At high viscosities, the dynamical torque only affects planetary embryos growing initially in the very outer regions of the disc. Cores forming in the inner region (a$<$5~au) of the disc are only barely affected. At low viscosites, the effect of the dynamical torque increases significantly. At $\alpha=0.0001$, if the dynamical torque is not accounted for, early forming planets have to originate outside of 30~au to stay exterior of 5~au, while planets can form as close as 15~au if the effects of the dynamical torque is included.

We show in Figure~\ref{fig:5} the result of planet formation simulations using $\alpha=0.0001$ for discs with higher aspect ratio ($h_0 =0.033$). The higher aspect ratio of the disc reduce the pebble accretion rates due to the higher scale height. The high aspect ratio also results in a higher pebble isolation mass resulting in larger planetary cores. This shortens the envelope contraction phase resulting in higher planetary masses, if the planets reach the pebble isolation mass and transition into gas accretion.

Our simulation results hold no surprise in this case, as a higher aspect ratio clearly results in higher planetary masses and at the same time also in faster inward migration. In addition it seems that the effects of the dynamical torque are less enhanced, because for a given planet mass the width of the horseshoe region decreases with increasing disc's aspect ratio.

Our results indicate that the effects of the CB accretion mechanism and the dynamical torque greatly reduce inward migration in low viscosity environments. As such these effects are important for future planet formation simulations.

\section{Conclusion} \label{Conclusions}

In this paper we have revisited the work of \cite{Bitsch-etal-2015} and \cite{Ndugu2018} on the accretion and migration of planets in a simple approach where pebble accretion is modeled with a constant Stokes number of 0.1.  \cite{Bitsch-etal-2015} concluded that giant planets that are at a few au from the central star at the end of the disc's lifetime must have started forming beyond 20~au. This conclusion appears inconsistent with the expectation that the most favorable place for the accretion of a giant planet's core should be the water ice line, which is probably not beyond $7$~au in young protoplanetary discs \citep{Savvidou2020arXiv200514097S}. This large-scale inward migration is driven by the high viscosity of the disc assumed in \cite{Bitsch-etal-2015}, leading to fast inward migration also in the type~II regime.

An obvious solution to reduce inward migration, is the reduction of the disc's viscosity. For low mid $\alpha$ viscosity, we have shown that the planet covers a smaller radial distance in type-II migration, as shown also by previous studies. In the case of low viscosities, the entropy driven corotation torque is fully saturated \citep[e.g.][]{Paardekooper2011}, preventing outward migration. Instead the low viscosity allows a transition into the slower type~II migration at lower planetary masses. The low viscosity of the disc limits also the gas accretion rate onto the planet, resulting in lower mass planets in discs with lower viscosity.

Our simulations show that the dynamical torque \citep{Paardekooper2014} reduces the inward migration before the planet opens a deep  gap in the disc. Furthermore, we find that accreting material directly from the horseshoe region \citep{Crida2017} can help in opening a deep gap, allowing an earlier transition into type~II migration and retaining the planet at larger distances from the star. 

 As already shown by \cite{Paardekooper2014}, a viscosity of below a few times $10^{-4}$ is needed for the dynamical torque to operate and slow down migration effectively. If the dynamical torque is as efficient as shown in our work, it could allow giant planets to form close to the water ice line and remain in the outer disc.  We thus conclude that the incorporation of the dynamical torques and the accretion from the horseshoe region can have significant influence on the resulting planet population and should be taken into account in future comprehensive planet population synthesis simulation.

\section*{Acknowledgments}
 N.N acknowledge financial support from the  International Science Program (ISP) and the  Poincar\'e junior fellowship. E.J appreciate the financial support from the  International Science Program (ISP). B.B, thanks the European Research Council (ERC Starting Grant 757448-PAMDORA) for their financial support. We thank an anonymous referee whose comments helped to improve this manuscript.
 
\section*{Data Availability}

The codes used to obtain the results in this paper is available upon reasonable request.

 \appendix

\section{Implementation of dynamical corotation torque}\label{dyntorque}
As hinted in subsection~\ref{planetform}, the dynamical corotation torque can effectively, reduce type~I migration rates in shallow discs with low disc viscosities. Due to this torque, the type~I migration rate of a planet is divided by a factor \citep[Equation~19 in][]{Paardekooper2014}. Therefore, the new type~I migration, $v_{\rm r}^{\rm I,dyn}$  in the limit of vanishing viscosity, becomes:
 \begin{equation}
  v_{\rm r}^{\rm I,dyn}  = \frac{v_{\rm r}^{\rm I}}{1-\frac{f(\cal{P})\cal{M}_{\rm HS}^{\rm loc}}{M_{\rm P}} + \frac{M_{\rm HS}}{M_{\rm p}}},
 \end{equation} where  $v_{\rm r}^{\rm I}$ is the classical type~I migration rate.\\ 
 $\mathcal{M}_{\rm HS}^{\rm loc} =2\pi a r_{\rm HS} {\Sigma}_{\rm disc}$ is the mass that the horseshoe region would have had if it had the unperturbed local density of the gas $\Sigma_{\rm disc}$; $M_{\rm HS}$ and $f(\cal{P})$ are computed as described in the subsection~\ref{gapaccimplement}. The factor $\left(1-\frac{f(\cal{P})\cal{M}_{\rm HS}^{\rm loc}}{M_{\rm p}} + \frac{M_{\rm HS}}{M_{\rm p}}\right)$ is typically larger than 1, which results in reduced migration. In fact, when the planet has not yet opened a gap by accretion and  gravitational pushing $\left(f_A\sim 1,~f({\cal{P}})\sim 1 \right)$, one has:
 \begin{eqnarray}
\label{dynext}
 {{f(\mathcal{P}) \mathcal{M}_{\rm HS}^{\rm loc}}\over{M_{\rm P}}}&\sim
&{{{\mathcal{M}_{\rm HS}^{\rm loc}}\over{\hat{M}_{\rm HS}}}=
{{\Sigma_{\rm disc}(a')}\over{\hat{\Sigma}_{\rm HS}(a')}}\sim
{{\Sigma_{\rm disc}(a)(a/a')^\beta}\over{\Sigma_{\rm disc}(a)(a/a')^{3/2}}}}\nonumber \\&&
={{(a/a')^{\beta-3/2}}},
\end{eqnarray} where $a'$ is the current location of the planet and $a$ is the location where 
the planet migrated from. The term $(a/a')^{\beta-3/2}$ is smaller than 1 if the
planet is migrating inward ($a'<a$) and the gradient of the surface density of 
the disc $\beta$ is smaller than $\frac{3}{2}$. Thus the term $-{f(\mathcal{P})} \mathcal{M}_{\rm HS}^{\rm loc} + M_{\rm HS} {>0}$, reducing the inward migration rate.

\section{Influence of Stokes number}\label{Stokes}

  \begin{figure}
        \centering
        \includegraphics[width=\columnwidth, 
height=2.45in]{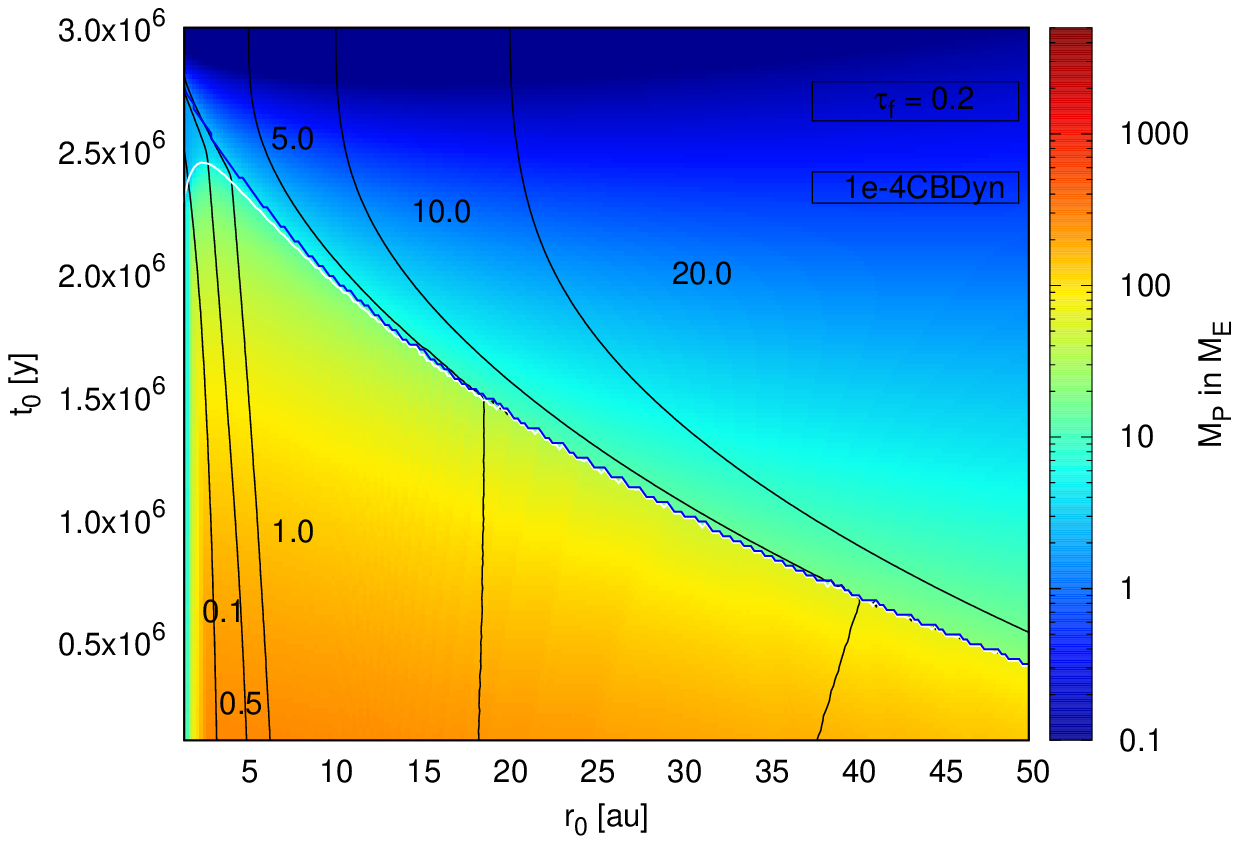} 
        \includegraphics[width=\columnwidth, height=2.45in]{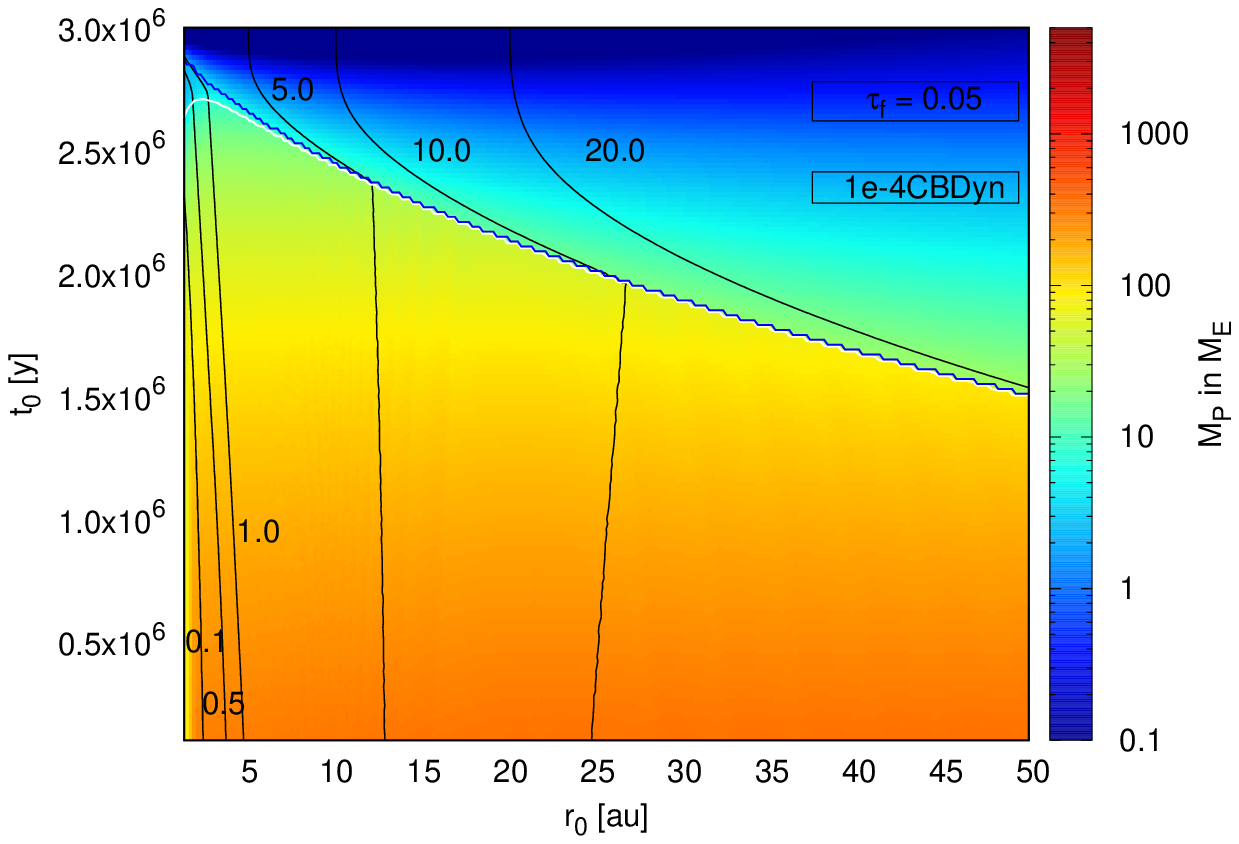}

  \caption{Growth maps showing the influence of different Stokes numbers of the pebbles for the growth of planets. We show the results of the 1e-4CBDyn model using a Stokes number of 0.2 (top) and 0.05 (bottom). The lines have the same meaning as in Figure~\ref{fig:2}.}
         \label{fig:6}
 \end{figure}
 
 In protoplanetary discs, the small micrometer dust grains can grow to form larger particles by coagulation \citep[e.g.][]{Brauer2008} or condensation \citep[e.g][]{RosJohansen2013}. These growth processes result in different Stokes numbers of particles, which are probably also not uniform in the disc. This has important consequences for the solid accretion rates via pebble accretion. For a fixed pebble surface density, larger Stokes number should lead to faster growth. However, in our model, the pebble surface density increases with reduction in Stokes number (Equation~\ref{eq:Sigmapebb}) for a fixed pebble flux. Therefore from Equation~\ref{eq:Mdot}, cores accrete pebbles at a rate proportional to ${\tau_{\rm}}^{\frac{-1}{3}}$, leading to faster growth with lower Stokes numbers.

We show the impact of different Stokes numbers in our model in Figure~\ref{fig:6}. Our results clearly indicate that planets growing in environments with larger/small Stokes numbers grow slower/faster and thus smaller/bigger. This can be clearly seen in Figure~\ref{fig:6}, where the simulations with lower Stokes numbers produce relatively massive giant planets compared to the simulations with larger Stokes numbers. On the other hand, it seems that the influence of Stokes number on the final position of the formed gas giants is not that pronounced at the inner parts of the disc (Figure~\ref{fig:6}).

We therefore conclude that the Stokes number has an important influence on the outcome of our simulations. However, the effects discussed in the main paper, namely the influence of the CB accretion mechanism and of the dynamical torques are untouched by the difference in Stokes number. In future simulations that aim to reproduce all planet populations from different Stokes numbers should be taken into account.

\bibliography{./Biblio.bib}

\begin{thebibliography}{}
\makeatletter
\relax
\def\mn@urlcharsother{\let\do\@makeother \do\$\do\&\do\#\do\^\do\_\do\%\do\~}
\def\mn@doi{\begingroup\mn@urlcharsother \@ifnextchar [ {\mn@doi@}
  {\mn@doi@[]}}
\def\mn@doi@[#1]#2{\def\@tempa{#1}\ifx\@tempa\@empty \href
  {http://dx.doi.org/#2} {doi:#2}\else \href {http://dx.doi.org/#2} {#1}\fi
  \endgroup}
\def\mn@eprint#1#2{\mn@eprint@#1:#2::\@nil}
\def\mn@eprint@arXiv#1{\href {http://arxiv.org/abs/#1} {{\tt arXiv:#1}}}
\def\mn@eprint@dblp#1{\href {http://dblp.uni-trier.de/rec/bibtex/#1.xml}
  {dblp:#1}}
\def\mn@eprint@#1:#2:#3:#4\@nil{\def\@tempa {#1}\def\@tempb {#2}\def\@tempc
  {#3}\ifx \@tempc \@empty \let \@tempc \@tempb \let \@tempb \@tempa \fi \ifx
  \@tempb \@empty \def\@tempb {arXiv}\fi \@ifundefined
  {mn@eprint@\@tempb}{\@tempb:\@tempc}{\expandafter \expandafter \csname
  mn@eprint@\@tempb\endcsname \expandafter{\@tempc}}}

\bibitem[\protect\citeauthoryear{{Ali-Dib}}{{Ali-Dib}}{2017}]{AliDib2017}
{Ali-Dib} M.,  2017, \mn@doi [\mnras] {10.1093/mnras/stx260}, \href
  {https://ui.adsabs.harvard.edu/abs/2017MNRAS.467.2845A} {467, 2845}

\bibitem[\protect\citeauthoryear{{Alibert}, {Mousis}, {Mordasini}  \&
  {Benz}}{{Alibert} et~al.}{2005}]{Alibert2005}
{Alibert} Y.,  {Mousis} O.,  {Mordasini} C.,   {Benz} W.,  2005, \mn@doi
  [\apjl] {10.1086/431325}, \href
  {https://ui.adsabs.harvard.edu/abs/2005ApJ...626L..57A} {626, L57}

\bibitem[\protect\citeauthoryear{{Alibert}, {Mordasini}  \& {Benz}}{{Alibert}
  et~al.}{2011}]{Alibert2011}
{Alibert} Y.,  {Mordasini} C.,   {Benz} W.,  2011, \mn@doi [\aap]
  {10.1051/0004-6361/201014760}, \href
  {https://ui.adsabs.harvard.edu/abs/2011A&A...526A..63A} {526, A63}

\bibitem[\protect\citeauthoryear{{Andrews} et~al.,}{{Andrews}
  et~al.}{2012}]{Andrews_2012}
{Andrews} S.~M.,  et~al., 2012, \mn@doi [\apj] {10.1088/0004-637X/744/2/162},
  \href {https://ui.adsabs.harvard.edu/abs/2012ApJ...744..162A} {744, 162}

\bibitem[\protect\citeauthoryear{{Ataiee}, {Baruteau}, {Alibert}  \&
  {Benz}}{{Ataiee} et~al.}{2018}]{Ataiee2018}
{Ataiee} S.,  {Baruteau} C.,  {Alibert} Y.,   {Benz} W.,  2018, \mn@doi [\aap]
  {10.1051/0004-6361/201732026}, \href
  {https://ui.adsabs.harvard.edu/abs/2018A&A...615A.110A} {615, A110}

\bibitem[\protect\citeauthoryear{{Baruteau} et~al.,}{{Baruteau}
  et~al.}{2014}]{Baruteau2014}
{Baruteau} C.,  et~al., 2014, \mn@doi [Protostars and Planets VI]
  {10.2458/azu_uapress_9780816531240-ch029}, \href
  {http://adsabs.harvard.edu/abs/2014prpl.conf..667B} {pp 667--689}

\bibitem[\protect\citeauthoryear{{Baumann} \& {Bitsch}}{{Baumann} \&
  {Bitsch}}{2020}]{Baumann2020}
{Baumann} T.,  {Bitsch} B.,  2020, arXiv e-prints, \href
  {https://ui.adsabs.harvard.edu/abs/2020arXiv200400874B} {p. arXiv:2004.00874}

\bibitem[\protect\citeauthoryear{{Ben{\'\i}tez-Llambay}, {Masset},
  {Koenigsberger}  \& {Szul{\'a}gyi}}{{Ben{\'\i}tez-Llambay}
  et~al.}{2015}]{Benitez2015}
{Ben{\'\i}tez-Llambay} P.,  {Masset} F.,  {Koenigsberger} G.,   {Szul{\'a}gyi}
  J.,  2015, \mn@doi [\nat] {10.1038/nature14277}, \href
  {https://ui.adsabs.harvard.edu/abs/2015Natur.520...63B} {520, 63}

\bibitem[\protect\citeauthoryear{{Bergez-Casalou}, {Bitsch}, {Pierens}, {Crida}
   \& {Raymond}}{{Bergez-Casalou} et~al.}{2020}]{Bergez2020arXiv201000485B}
{Bergez-Casalou} C.,  {Bitsch} B.,  {Pierens} A.,  {Crida} A.,   {Raymond}
  S.~N.,  2020, arXiv e-prints, \href
  {https://ui.adsabs.harvard.edu/abs/2020arXiv201000485B} {p. arXiv:2010.00485}

\bibitem[\protect\citeauthoryear{{Bitsch} \& {Johansen}}{{Bitsch} \&
  {Johansen}}{2017}]{BitschJohansen2017}
{Bitsch} B.,  {Johansen} A.,  2017, in {Pessah} M.,  {Gressel} O.,  eds,
  Astrophysics and Space Science Library Vol. 445, Astrophysics and Space
  Science Library. p.~339, \mn@doi{10.1007/978-3-319-60609-5_12}

\bibitem[\protect\citeauthoryear{{Bitsch}, {Morbidelli}, {Lega}  \&
  {Crida}}{{Bitsch} et~al.}{2014}]{Bitsch2014}
{Bitsch} B.,  {Morbidelli} A.,  {Lega} E.,   {Crida} A.,  2014, \mn@doi [\aap]
  {10.1051/0004-6361}, \href
  {http://adsabs.harvard.edu/abs/2014A\%26A...564A.135B} {564, A135}

\bibitem[\protect\citeauthoryear{{Bitsch}, {Johansen}, {Lambrechts}  \&
  {Morbidelli}}{{Bitsch} et~al.}{2015a}]{Bitsch}
{Bitsch} B.,  {Johansen} A.,  {Lambrechts} M.,   {Morbidelli} A.,  2015a,
  \mn@doi [\aap] {10.1051/0004-6361}, \href
  {http://adsabs.harvard.edu/abs/2015A\%26A...575A..28B} {575, A28}

\bibitem[\protect\citeauthoryear{{Bitsch}, {Lambrechts}  \&
  {Johansen}}{{Bitsch} et~al.}{2015b}]{Bitsch-etal-2015}
{Bitsch} B.,  {Lambrechts} M.,   {Johansen} A.,  2015b, \mn@doi [\aap]
  {10.1051/0004-6361}, \href
  {http://adsabs.harvard.edu/abs/2015A\%26A...582A.112B} {582, A112}

\bibitem[\protect\citeauthoryear{{Bitsch}, {Morbidelli}, {Johansen}, {Lega},
  {Lambrechts}  \& {Crida}}{{Bitsch} et~al.}{2018}]{Bitsch2018}
{Bitsch} B.,  {Morbidelli} A.,  {Johansen} A.,  {Lega} E.,  {Lambrechts} M.,
  {Crida} A.,  2018, \mn@doi [\aap] {10.1051/0004-6361/201731931}, \href
  {https://ui.adsabs.harvard.edu/abs/2018A&A...612A..30B} {612, A30}

\bibitem[\protect\citeauthoryear{{Bottke}, {Durda}, {Nesvorn{\'y}}, {Jedicke},
  {Morbidelli}, {Vokrouhlick{\'y}}  \& {Levison}}{{Bottke}
  et~al.}{2005}]{Bottke2005}
{Bottke} W.~F.,  {Durda} D.~D.,  {Nesvorn{\'y}} D.,  {Jedicke} R.,
  {Morbidelli} A.,  {Vokrouhlick{\'y}} D.,   {Levison} H.,  2005, \mn@doi
  [\icarus] {10.1016/j.icarus.2004.10.026}, \href
  {https://ui.adsabs.harvard.edu/abs/2005Icar..175..111B} {175, 111}

\bibitem[\protect\citeauthoryear{{Brauer}, {Dullemond}  \& {Henning}}{{Brauer}
  et~al.}{2008}]{Brauer2008}
{Brauer} F.,  {Dullemond} C.~P.,   {Henning} T.,  2008, \mn@doi [\aap]
  {10.1051/0004-6361:20077759}, \href
  {https://ui.adsabs.harvard.edu/abs/2008A&A...480..859B} {480, 859}

\bibitem[\protect\citeauthoryear{{Br{\"u}gger}, {Alibert}, {Ataiee}  \&
  {Benz}}{{Br{\"u}gger} et~al.}{2018}]{Brugger2018}
{Br{\"u}gger} N.,  {Alibert} Y.,  {Ataiee} S.,   {Benz} W.,  2018, \mn@doi
  [\aap] {10.1051/0004-6361/201833347}, \href
  {http://adsabs.harvard.edu/abs/2018A%26A...619A.174B} {619, A174}

\bibitem[\protect\citeauthoryear{{Cassan} et~al.,}{{Cassan}
  et~al.}{2012}]{Cassan2012}
{Cassan} A.,  et~al., 2012, \mn@doi [\nat] {10.1038/nature10684}, \href
  {https://ui.adsabs.harvard.edu/abs/2012Natur.481..167C} {481, 167}

\bibitem[\protect\citeauthoryear{{Chambers}}{{Chambers}}{2018}]{Chambers2018}
{Chambers} J.,  2018, \mn@doi [\apj] {10.3847/1538-4357/aada09}, \href
  {http://adsabs.harvard.edu/abs/2018ApJ...865...30C} {865, 30}

\bibitem[\protect\citeauthoryear{{Chiang} \& {Goldreich}}{{Chiang} \&
  {Goldreich}}{1997}]{Chiang1997ApJ...490..368C}
{Chiang} E.~I.,  {Goldreich} P.,  1997, \mn@doi [\apj] {10.1086/304869}, \href
  {https://ui.adsabs.harvard.edu/abs/1997ApJ...490..368C} {490, 368}

\bibitem[\protect\citeauthoryear{{Crida} \& {Bitsch}}{{Crida} \&
  {Bitsch}}{2017}]{Crida2017}
{Crida} A.,  {Bitsch} B.,  2017, \mn@doi [\icarus]
  {10.1016/j.icarus.2016.10.017}, \href
  {http://adsabs.harvard.edu/abs/2017Icar..285..145C} {285, 145}

\bibitem[\protect\citeauthoryear{{Crida} \& {Morbidelli}}{{Crida} \&
  {Morbidelli}}{2007}]{Crida2007}
{Crida} A.,  {Morbidelli} A.,  2007, \mn@doi [\mnras]
  {10.1111/j.1365-2966.2007.11704.x}, \href
  {http://adsabs.harvard.edu/abs/2007MNRAS.377.1324C} {377, 1324}

\bibitem[\protect\citeauthoryear{{Crida}, {Morbidelli}  \& {Masset}}{{Crida}
  et~al.}{2006}]{Crida2006}
{Crida} A.,  {Morbidelli} A.,   {Masset} F.,  2006, \mn@doi [\icarus]
  {10.1016/j.icarus.2005.10.007}, \href
  {http://adsabs.harvard.edu/abs/2006Icar..181..587C} {181, 587}

\bibitem[\protect\citeauthoryear{{Cridland}, {Eistrup}  \& {van
  Dishoeck}}{{Cridland} et~al.}{2019}]{Cridland2019}
{Cridland} A.~J.,  {Eistrup} C.,   {van Dishoeck} E.~F.,  2019, \mn@doi [\aap]
  {10.1051/0004-6361/201834378}, \href
  {https://ui.adsabs.harvard.edu/abs/2019A&A...627A.127C} {627, A127}

\bibitem[\protect\citeauthoryear{{Dittkrist}, {Mordasini}, {Klahr}, {Alibert}
  \& {Henning}}{{Dittkrist} et~al.}{2014}]{Dittkrist2014}
{Dittkrist} K.-M.,  {Mordasini} C.,  {Klahr} H.,  {Alibert} Y.,   {Henning} T.,
   2014, \mn@doi [\aap] {10.1051/0004-6361}, \href
  {http://adsabs.harvard.edu/abs/2014A\%26A...567A.121D} {567, A121}

\bibitem[\protect\citeauthoryear{{Dr{\k{a}}{\.z}kowska} \&
  {Alibert}}{{Dr{\k{a}}{\.z}kowska} \&
  {Alibert}}{2017}]{Drazkowska2017A&A...608A..92D}
{Dr{\k{a}}{\.z}kowska} J.,  {Alibert} Y.,  2017, \mn@doi [\aap]
  {10.1051/0004-6361/201731491}, \href
  {https://ui.adsabs.harvard.edu/abs/2017A&A...608A..92D} {608, A92}

\bibitem[\protect\citeauthoryear{{Fischer} \& {Valenti}}{{Fischer} \&
  {Valenti}}{2005}]{Fischer2005}
{Fischer} D.~A.,  {Valenti} J.,  2005, \mn@doi [\apj] {10.1086/428383}, \href
  {https://ui.adsabs.harvard.edu/abs/2005ApJ...622.1102F} {622, 1102}

\bibitem[\protect\citeauthoryear{{Flock}, {Turner}, {Mulders}, {Hasegawa},
  {Nelson}  \& {Bitsch}}{{Flock} et~al.}{2019}]{Flock2019}
{Flock} M.,  {Turner} N.~J.,  {Mulders} G.~D.,  {Hasegawa} Y.,  {Nelson} R.~P.,
    {Bitsch} B.,  2019, \mn@doi [\aap] {10.1051/0004-6361/201935806}, \href
  {https://ui.adsabs.harvard.edu/abs/2019A&A...630A.147F} {630, A147}

\bibitem[\protect\citeauthoryear{{Fortier}, {Alibert}, {Carron}, {Benz}  \&
  {Dittkrist}}{{Fortier} et~al.}{2013}]{Fortier2013}
{Fortier} A.,  {Alibert} Y.,  {Carron} F.,  {Benz} W.,   {Dittkrist} K.~M.,
  2013, \mn@doi [\aap] {10.1051/0004-6361/201220241}, \href
  {https://ui.adsabs.harvard.edu/abs/2013A&A...549A..44F} {549, A44}

\bibitem[\protect\citeauthoryear{{Fressin} et~al.,}{{Fressin}
  et~al.}{2013}]{Fressin2013}
{Fressin} F.,  et~al., 2013, \mn@doi [\apj] {10.1088/0004-637X/766/2/81}, \href
  {https://ui.adsabs.harvard.edu/abs/2013ApJ...766...81F} {766, 81}

\bibitem[\protect\citeauthoryear{{Haisch}, {Lada}  \& {Lada}}{{Haisch}
  et~al.}{2001}]{Haisch_2001}
{Haisch} Karl~E. J.,  {Lada} E.~A.,   {Lada} C.~J.,  2001, \mn@doi [\apjl]
  {10.1086/320685}, \href
  {https://ui.adsabs.harvard.edu/abs/2001ApJ...553L.153H} {553, L153}

\bibitem[\protect\citeauthoryear{{Hartmann}, {Calvet}, {Gullbring}  \&
  {D'Alessio}}{{Hartmann} et~al.}{1998}]{Hartmann1998}
{Hartmann} L.,  {Calvet} N.,  {Gullbring} E.,   {D'Alessio} P.,  1998, \mn@doi
  [\apj] {10.1086/305277}, \href
  {http://adsabs.harvard.edu/abs/1998ApJ...495..385H} {495, 385}

\bibitem[\protect\citeauthoryear{{Howard} et~al.,}{{Howard}
  et~al.}{2010}]{Howard2010}
{Howard} A.~W.,  et~al., 2010, \mn@doi [Science] {10.1126/science.1194854},
  \href {https://ui.adsabs.harvard.edu/abs/2010Sci...330..653H} {330, 653}

\bibitem[\protect\citeauthoryear{{Ida} \& {Guillot}}{{Ida} \&
  {Guillot}}{2016}]{IdaGuillot2016}
{Ida} S.,  {Guillot} T.,  2016, \mn@doi [\aap] {10.1051/0004-6361}, \href
  {http://adsabs.harvard.edu/abs/2016A\%26A...596L...3I} {596, L3}

\bibitem[\protect\citeauthoryear{{Ida} \& {Lin}}{{Ida} \&
  {Lin}}{2004}]{Ida2004}
{Ida} S.,  {Lin} D.~N.~C.,  2004, \mn@doi [\apj] {10.1086/381724}, \href
  {http://adsabs.harvard.edu/abs/2004ApJ...604..388I} {604, 388}

\bibitem[\protect\citeauthoryear{{Ida} \& {Lin}}{{Ida} \&
  {Lin}}{2008a}]{Ida2008a}
{Ida} S.,  {Lin} D.~N.~C.,  2008a, \mn@doi [\apj] {10.1086/523754}, \href
  {https://ui.adsabs.harvard.edu/abs/2008ApJ...673..487I} {673, 487}

\bibitem[\protect\citeauthoryear{{Ida} \& {Lin}}{{Ida} \&
  {Lin}}{2008b}]{Ida2008b}
{Ida} S.,  {Lin} D.~N.~C.,  2008b, \mn@doi [\apj] {10.1086/590401}, \href
  {https://ui.adsabs.harvard.edu/abs/2008ApJ...685..584I} {685, 584}

\bibitem[\protect\citeauthoryear{{Ida}, {Lin}  \& {Nagasawa}}{{Ida}
  et~al.}{2013}]{Ida2013}
{Ida} S.,  {Lin} D.~N.~C.,   {Nagasawa} M.,  2013, \mn@doi [\apj]
  {10.1088/0004-637X/775/1/42}, \href
  {https://ui.adsabs.harvard.edu/abs/2013ApJ...775...42I} {775, 42}

\bibitem[\protect\citeauthoryear{{Ida}, {Tanaka}, {Johansen}, {Kanagawa}  \&
  {Tanigawa}}{{Ida} et~al.}{2018}]{Ida2018}
{Ida} S.,  {Tanaka} H.,  {Johansen} A.,  {Kanagawa} K.~D.,   {Tanigawa} T.,
  2018, \mn@doi [\apj] {10.3847/1538-4357/aad69c}, \href
  {https://ui.adsabs.harvard.edu/abs/2018ApJ...864...77I} {864, 77}

\bibitem[\protect\citeauthoryear{{Ikoma}, {Nakazawa}  \& {Emori}}{{Ikoma}
  et~al.}{2000}]{Ikoma2000}
{Ikoma} M.,  {Nakazawa} K.,   {Emori} H.,  2000, \mn@doi [\apj]
  {10.1086/309050}, \href
  {https://ui.adsabs.harvard.edu/abs/2000ApJ...537.1013I} {537, 1013}

\bibitem[\protect\citeauthoryear{{Johansen} \& {Bitsch}}{{Johansen} \&
  {Bitsch}}{2019}]{JohansenBitsch2019}
{Johansen} A.,  {Bitsch} B.,  2019, arXiv e-prints, \href
  {https://ui.adsabs.harvard.edu/abs/2019arXiv190910429J} {p. arXiv:1909.10429}

\bibitem[\protect\citeauthoryear{{Johansen} \& {Lacerda}}{{Johansen} \&
  {Lacerda}}{2010}]{JohansenLarcda2010}
{Johansen} A.,  {Lacerda} P.,  2010, \mn@doi [\mnras]
  {10.1111/j.1365-2966.2010.16309.x}, \href
  {https://ui.adsabs.harvard.edu/abs/2010MNRAS.404..475J} {404, 475}

\bibitem[\protect\citeauthoryear{{Johansen}, {Ida}  \& {Brasser}}{{Johansen}
  et~al.}{2019}]{Johansen2019}
{Johansen} A.,  {Ida} S.,   {Brasser} R.,  2019, \mn@doi [\aap]
  {10.1051/0004-6361/201834071}, \href
  {https://ui.adsabs.harvard.edu/abs/2019A&A...622A.202J} {622, A202}

\bibitem[\protect\citeauthoryear{{Johnson}, {Aller}, {Howard}  \&
  {Crepp}}{{Johnson} et~al.}{2010}]{Johnson2010}
{Johnson} J.~A.,  {Aller} K.~M.,  {Howard} A.~W.,   {Crepp} J.~R.,  2010,
  \mn@doi [\pasp] {10.1086/655775}, \href
  {http://adsabs.harvard.edu/abs/2010PASP..122..905J} {122, 905}

\bibitem[\protect\citeauthoryear{{Kanagawa}, {Tanaka}  \&
  {Szuszkiewicz}}{{Kanagawa} et~al.}{2018}]{Kanagawa2018}
{Kanagawa} K.~D.,  {Tanaka} H.,   {Szuszkiewicz} E.,  2018, \mn@doi [\apj]
  {10.3847/1538-4357/aac8d9}, \href
  {http://adsabs.harvard.edu/abs/2018ApJ...861..140K} {861, 140}

\bibitem[\protect\citeauthoryear{{Lambrechts} \& {Johansen}}{{Lambrechts} \&
  {Johansen}}{2012}]{Lambrechts2012}
{Lambrechts} M.,  {Johansen} A.,  2012, \mn@doi [\aap] {10.1051/0004-6361},
  \href {http://adsabs.harvard.edu/abs/2012A\%26A...544A..32L} {544, A32}

\bibitem[\protect\citeauthoryear{{Lambrechts} \& {Johansen}}{{Lambrechts} \&
  {Johansen}}{2014}]{Lambrechts2014}
{Lambrechts} M.,  {Johansen} A.,  2014, \mn@doi [\aap] {10.1051/0004-6361},
  \href {http://adsabs.harvard.edu/abs/2014A\%26A...572A.107L} {572, A107}

\bibitem[\protect\citeauthoryear{{Lambrechts}, {Johansen}  \&
  {Morbidelli}}{{Lambrechts} et~al.}{2014}]{Lambrechts2014b}
{Lambrechts} M.,  {Johansen} A.,   {Morbidelli} A.,  2014, \mn@doi [\aap]
  {10.1051/0004-6361}, \href
  {http://adsabs.harvard.edu/abs/2014A\%26A...572A..35L} {572, A35}

\bibitem[\protect\citeauthoryear{{Laughlin} \& {Adams}}{{Laughlin} \&
  {Adams}}{1997}]{Laughlin1997}
{Laughlin} G.,  {Adams} F.~C.,  1997, \mn@doi [\apjl] {10.1086/311056}, \href
  {http://adsabs.harvard.edu/abs/1997ApJ...491L..51L} {491, L51}

\bibitem[\protect\citeauthoryear{{Levison}, {Thommes}  \& {Duncan}}{{Levison}
  et~al.}{2010}]{Levison2010}
{Levison} H.~F.,  {Thommes} E.,   {Duncan} M.~J.,  2010, \mn@doi [\aj]
  {10.1088/0004-6256}, \href
  {http://adsabs.harvard.edu/abs/2010AJ....139.1297L} {139, 1297}

\bibitem[\protect\citeauthoryear{{Lin} \& {Ida}}{{Lin} \&
  {Ida}}{1997}]{Lin1997}
{Lin} D.~N.~C.,  {Ida} S.,  1997, \mn@doi [\apj] {10.1086/303738}, \href
  {http://adsabs.harvard.edu/abs/1997ApJ...477..781L} {477, 781}

\bibitem[\protect\citeauthoryear{{Lin} \& {Papaloizou}}{{Lin} \&
  {Papaloizou}}{1986}]{Lin1986a}
{Lin} D.~N.~C.,  {Papaloizou} J.,  1986, \mn@doi [\apj] {10.1086/164426}, \href
  {https://ui.adsabs.harvard.edu/abs/1986ApJ...307..395L} {307, 395}

\bibitem[\protect\citeauthoryear{{Lubow} \& {D'Angelo}}{{Lubow} \&
  {D'Angelo}}{2006}]{Lubow2006}
{Lubow} S.~H.,  {D'Angelo} G.,  2006, \mn@doi [\apj] {10.1086/500356}, \href
  {https://ui.adsabs.harvard.edu/abs/2006ApJ...641..526L} {641, 526}

\bibitem[\protect\citeauthoryear{{Lynden-Bell} \& {Pringle}}{{Lynden-Bell} \&
  {Pringle}}{1974}]{Lynden-Bell1974}
{Lynden-Bell} D.,  {Pringle} J.~E.,  1974, \mn@doi [\mnras]
  {10.1093/mnras/168.3.603}, \href
  {https://ui.adsabs.harvard.edu/abs/1974MNRAS.168..603L} {168, 603}

\bibitem[\protect\citeauthoryear{{Machida}, {Kokubo}, {Inutsuka}  \&
  {Matsumoto}}{{Machida} et~al.}{2010}]{Machida2010}
{Machida} M.~N.,  {Kokubo} E.,  {Inutsuka} S.-I.,   {Matsumoto} T.,  2010,
  \mn@doi [\mnras] {10.1111/j.1365-2966.2010.16527.x}, \href
  {http://adsabs.harvard.edu/abs/2010MNRAS.405.1227M} {405, 1227}

\bibitem[\protect\citeauthoryear{{Manara}, {Robberto}, {Da Rio}, {Lodato},
  {Hillenbrand}, {Stassun}  \& {Soderblom}}{{Manara} et~al.}{2012}]{Manara2012}
{Manara} C.~F.,  {Robberto} M.,  {Da Rio} N.,  {Lodato} G.,  {Hillenbrand}
  L.~A.,  {Stassun} K.~G.,   {Soderblom} D.~R.,  2012, \mn@doi [\apj]
  {10.1088/0004-637X}, \href
  {http://adsabs.harvard.edu/abs/2012ApJ...755..154M} {755, 154}

\bibitem[\protect\citeauthoryear{{Masset}, {Morbidelli}, {Crida}  \&
  {Ferreira}}{{Masset} et~al.}{2006}]{Masset2006}
{Masset} F.~S.,  {Morbidelli} A.,  {Crida} A.,   {Ferreira} J.,  2006, \mn@doi
  [\apj] {10.1086/500967}, \href
  {https://ui.adsabs.harvard.edu/abs/2006ApJ...642..478M} {642, 478}

\bibitem[\protect\citeauthoryear{{Mayor} et~al.,}{{Mayor}
  et~al.}{1995}]{Mayor1995}
{Mayor} M.,  et~al., 1995, \iaucirc, \href
  {https://ui.adsabs.harvard.edu/abs/1995IAUC.6251....1M} {6251, 1}

\bibitem[\protect\citeauthoryear{{McNeil}, {Duncan}  \& {Levison}}{{McNeil}
  et~al.}{2005}]{McNeil_2005}
{McNeil} D.,  {Duncan} M.,   {Levison} H.~F.,  2005, \mn@doi [\aj]
  {10.1086/497687}, \href
  {https://ui.adsabs.harvard.edu/abs/2005AJ....130.2884M} {130, 2884}

\bibitem[\protect\citeauthoryear{{Miguel}, {Guilera}  \& {Brunini}}{{Miguel}
  et~al.}{2011}]{Miguel2011}
{Miguel} Y.,  {Guilera} O.~M.,   {Brunini} A.,  2011, \mn@doi [\mnras]
  {10.1111/j.1365-2966.2010.17887.x}, \href
  {https://ui.adsabs.harvard.edu/abs/2011MNRAS.412.2113M} {412, 2113}

\bibitem[\protect\citeauthoryear{{Morales} et~al.,}{{Morales}
  et~al.}{2019}]{Morales2019}
{Morales} J.~C.,  et~al., 2019, arXiv e-prints, \href
  {https://ui.adsabs.harvard.edu/abs/2019arXiv190912174M} {p. arXiv:1909.12174}

\bibitem[\protect\citeauthoryear{{Morbidelli}}{{Morbidelli}}{2015}]{Morbidelli2015}
{Morbidelli} A.,  2015, European Planetary Science Congress 2015, held 27
  September - 2 October, 2015 in Nantes, France, Online at
  http://meetingorganizer.copernicus.org/EPSC2015, id.EPSC2015-30, \href
  {http://adsabs.harvard.edu/abs/2015EPSC...10...30M} {10, EPSC2015}

\bibitem[\protect\citeauthoryear{{Morbidelli}, {Bottke}, {Nesvorn{\'y}}  \&
  {Levison}}{{Morbidelli} et~al.}{2009}]{Morbidelli2009}
{Morbidelli} A.,  {Bottke} W.~F.,  {Nesvorn{\'y}} D.,   {Levison} H.~F.,  2009,
  \mn@doi [\icarus] {10.1016/j.icarus.2009.07.011}, \href
  {https://ui.adsabs.harvard.edu/abs/2009Icar..204..558M} {204, 558}

\bibitem[\protect\citeauthoryear{{Mordasini}, {Alibert}  \& {Benz}}{{Mordasini}
  et~al.}{2009}]{Mordasini2009}
{Mordasini} C.,  {Alibert} Y.,   {Benz} W.,  2009, \mn@doi [\aap]
  {10.1051/0004-6361}, \href
  {http://adsabs.harvard.edu/abs/2009A\%26A...501.1139M} {501, 1139}

\bibitem[\protect\citeauthoryear{{Mordasini}, {Molli{\`e}re}, {Dittkrist},
  {Jin}  \& {Alibert}}{{Mordasini} et~al.}{2015}]{Mordasini2015}
{Mordasini} C.,  {Molli{\`e}re} P.,  {Dittkrist} K.~M.,  {Jin} S.,   {Alibert}
  Y.,  2015, \mn@doi [International Journal of Astrobiology]
  {10.1017/S1473550414000263}, \href
  {https://ui.adsabs.harvard.edu/abs/2015IJAsB..14..201M} {14, 201}

\bibitem[\protect\citeauthoryear{{Movshovitz} \& {Podolak}}{{Movshovitz} \&
  {Podolak}}{2008}]{Movshovitz2008}
{Movshovitz} N.,  {Podolak} M.,  2008, \mn@doi [\icarus]
  {10.1016/j.icarus.2007.09.018}, \href
  {https://ui.adsabs.harvard.edu/abs/2008Icar..194..368M} {194, 368}

\bibitem[\protect\citeauthoryear{{Nayakshin}}{{Nayakshin}}{2015}]{Nayakshin2015MNRAS.454...64N}
{Nayakshin} S.,  2015, \mn@doi [\mnras] {10.1093/mnras/stv1915}, \href
  {https://ui.adsabs.harvard.edu/abs/2015MNRAS.454...64N} {454, 64}

\bibitem[\protect\citeauthoryear{{Ndugu}, {Bitsch}  \& {Jurua}}{{Ndugu}
  et~al.}{2018}]{Ndugu2018}
{Ndugu} N.,  {Bitsch} B.,   {Jurua} E.,  2018, \mn@doi [\mnras]
  {10.1093/mnras/stx2815}, \href
  {http://adsabs.harvard.edu/abs/2018MNRAS.474..886N} {474, 886}

\bibitem[\protect\citeauthoryear{{Ndugu}, {Bitsch}  \& {Jurua}}{{Ndugu}
  et~al.}{2019}]{Ndugu2019}
{Ndugu} N.,  {Bitsch} B.,   {Jurua} E.,  2019, \mn@doi [\mnras]
  {10.1093/mnras/stz1862}, \href
  {https://ui.adsabs.harvard.edu/abs/2019MNRAS.488.3625N} {488, 3625}

\bibitem[\protect\citeauthoryear{{Nelson}, {Papaloizou}, {Masset}  \&
  {Kley}}{{Nelson} et~al.}{2000}]{Nelson2000}
{Nelson} R.~P.,  {Papaloizou} J.~C.~B.,  {Masset} F.,   {Kley} W.,  2000,
  \mn@doi [\mnras] {10.1046/j.1365-8711.2000.03605.x}, \href
  {http://adsabs.harvard.edu/abs/2000MNRAS.318...18N} {318, 18}

\bibitem[\protect\citeauthoryear{{Nelson}, {Gressel}  \& {Umurhan}}{{Nelson}
  et~al.}{2013}]{Nelson2013}
{Nelson} R.~P.,  {Gressel} O.,   {Umurhan} O.~M.,  2013, \mn@doi [\mnras]
  {10.1093/mnras}, \href {http://adsabs.harvard.edu/abs/2013MNRAS.435.2610N}
  {435, 2610}

\bibitem[\protect\citeauthoryear{{Ogihara}, {Morbidelli}  \&
  {Guillot}}{{Ogihara} et~al.}{2015}]{Ogihara2015}
{Ogihara} M.,  {Morbidelli} A.,   {Guillot} T.,  2015, \mn@doi [\aap]
  {10.1051/0004-6361}, \href
  {http://adsabs.harvard.edu/abs/2015A\%26A...578A..36O} {578, A36}

\bibitem[\protect\citeauthoryear{{Ormel} \& {Klahr}}{{Ormel} \&
  {Klahr}}{2010}]{OrmelKlahr2010}
{Ormel} C.~W.,  {Klahr} H.~H.,  2010, \mn@doi [\aap]
  {10.1051/0004-6361/201014903}, \href
  {https://ui.adsabs.harvard.edu/abs/2010A&A...520A..43O} {520, A43}

\bibitem[\protect\citeauthoryear{{Paardekooper}}{{Paardekooper}}{2014}]{Paardekooper2014}
{Paardekooper} S.-J.,  2014, \mn@doi [\mnras] {10.1093/mnras}, \href
  {http://adsabs.harvard.edu/abs/2014MNRAS.444.2031P} {444, 2031}

\bibitem[\protect\citeauthoryear{{Paardekooper}, {Baruteau}  \&
  {Kley}}{{Paardekooper} et~al.}{2011}]{Paardekooper2011}
{Paardekooper} S.-J.,  {Baruteau} C.,   {Kley} W.,  2011, \mn@doi [\mnras]
  {10.1111/j.1365-2966.2010.17442.x}, \href
  {http://adsabs.harvard.edu/abs/2011MNRAS.410..293P} {410, 293}

\bibitem[\protect\citeauthoryear{{Piso} \& {Youdin}}{{Piso} \&
  {Youdin}}{2014}]{Pisso2014}
{Piso} A.-M.~A.,  {Youdin} A.~N.,  2014, \mn@doi [\apj] {10.1088/0004-637X},
  \href {http://adsabs.harvard.edu/abs/2014ApJ...786...21P} {786, 21}

\bibitem[\protect\citeauthoryear{{Pollack}, {Hubickyj}, {Bodenheimer},
  {Lissauer}, {Podolak}  \& {Greenzweig}}{{Pollack} et~al.}{1996}]{Pollack1996}
{Pollack} J.~B.,  {Hubickyj} O.,  {Bodenheimer} P.,  {Lissauer} J.~J.,
  {Podolak} M.,   {Greenzweig} Y.,  1996, \mn@doi [\icarus]
  {10.1006/icar.1996.0190}, \href
  {http://adsabs.harvard.edu/abs/1996Icar..124...62P} {124, 62}

\bibitem[\protect\citeauthoryear{{Ronco}, {Guilera}  \& {de El{\'\i}a}}{{Ronco}
  et~al.}{2017}]{Ronco2017}
{Ronco} M.~P.,  {Guilera} O.~M.,   {de El{\'\i}a} G.~C.,  2017, \mn@doi
  [\mnras] {10.1093/mnras/stx1746}, \href
  {https://ui.adsabs.harvard.edu/abs/2017MNRAS.471.2753R} {471, 2753}

\bibitem[\protect\citeauthoryear{{Ros} \& {Johansen}}{{Ros} \&
  {Johansen}}{2013}]{RosJohansen2013}
{Ros} K.,  {Johansen} A.,  2013, \mn@doi [\aap] {10.1051/0004-6361/201220536},
  \href {https://ui.adsabs.harvard.edu/abs/2013A&A...552A.137R} {552, A137}

\bibitem[\protect\citeauthoryear{{Santos}, {Mayor}, {Naef}, {Pepe}, {Queloz}
  \& {Udry}}{{Santos} et~al.}{2004}]{Santos2004}
{Santos} N.~C.,  {Mayor} M.,  {Naef} D.,  {Pepe} F.,  {Queloz} D.,   {Udry} S.,
   2004, in {Dupree} A.~K.,  {Benz} A.~O.,  eds,  IAU Symposium Vol. 219, Stars
  as Suns : Activity, Evolution and Planets. p.~311

\bibitem[\protect\citeauthoryear{{Savvidou}, {Bitsch}  \&
  {Lambrechts}}{{Savvidou} et~al.}{2020}]{Savvidou2020arXiv200514097S}
{Savvidou} S.,  {Bitsch} B.,   {Lambrechts} M.,  2020, arXiv e-prints, \href
  {https://ui.adsabs.harvard.edu/abs/2020arXiv200514097S} {p. arXiv:2005.14097}

\bibitem[\protect\citeauthoryear{{Schoonenberg} \& {Ormel}}{{Schoonenberg} \&
  {Ormel}}{2017}]{Schoonenberg2017}
{Schoonenberg} D.,  {Ormel} C.~W.,  2017, preprint, \href
  {http://adsabs.harvard.edu/abs/2017arXiv170202151S} {} (\mn@eprint {arXiv}
  {1702.02151})

\bibitem[\protect\citeauthoryear{{Shakura} \& {Sunyaev}}{{Shakura} \&
  {Sunyaev}}{1973}]{Shakura1973}
{Shakura} N.~I.,  {Sunyaev} R.~A.,  1973, \aap, \href
  {http://adsabs.harvard.edu/abs/1973A\%26A....24..337S} {24, 337}

\bibitem[\protect\citeauthoryear{{Singer} et~al.,}{{Singer}
  et~al.}{2019}]{Singer2019}
{Singer} K.~N.,  et~al., 2019, \mn@doi [Science] {10.1126/science.aap8628},
  \href {https://ui.adsabs.harvard.edu/abs/2019Sci...363..955S} {363, 955}

\bibitem[\protect\citeauthoryear{{Stoll} \& {Kley}}{{Stoll} \&
  {Kley}}{2014}]{Stoll2014}
{Stoll} M.~H.~R.,  {Kley} W.,  2014, \mn@doi [\aap] {10.1051/0004-6361}, \href
  {http://adsabs.harvard.edu/abs/2014A\%26A...572A..77S} {572, A77}

\bibitem[\protect\citeauthoryear{{Suzuki} et~al.,}{{Suzuki}
  et~al.}{2016}]{Suzuki2016}
{Suzuki} D.,  et~al., 2016, \mn@doi [\apj] {10.3847/1538-4357/833/2/145}, \href
  {https://ui.adsabs.harvard.edu/abs/2016ApJ...833..145S} {833, 145}

\bibitem[\protect\citeauthoryear{{Tanaka} \& {Ida}}{{Tanaka} \&
  {Ida}}{1999}]{TanakaIda1999}
{Tanaka} H.,  {Ida} S.,  1999, \mn@doi [\icarus] {10.1006/icar.1999.6107},
  \href {https://ui.adsabs.harvard.edu/abs/1999Icar..139..350T} {139, 350}

\bibitem[\protect\citeauthoryear{{Thommes}, {Duncan}  \& {Levison}}{{Thommes}
  et~al.}{2003}]{Thommes2003}
{Thommes} E.~W.,  {Duncan} M.~J.,   {Levison} H.~F.,  2003, \mn@doi [\icarus]
  {10.1016/S0019-1035(02)00043-X}, \href
  {https://ui.adsabs.harvard.edu/abs/2003Icar..161..431T} {161, 431}

\bibitem[\protect\citeauthoryear{{Turner}, {Fromang}, {Gammie}, {Klahr},
  {Lesur}, {Wardle}  \& {Bai}}{{Turner} et~al.}{2014}]{Turner2014}
{Turner} N.~J.,  {Fromang} S.,  {Gammie} C.,  {Klahr} H.,  {Lesur} G.,
  {Wardle} M.,   {Bai} X.-N.,  2014, \mn@doi [Protostars and Planets VI]
  {10.2458/azu_uapress_9780816531240-ch018}, \href
  {http://adsabs.harvard.edu/abs/2014prpl.conf..411T} {pp 411--432}

\bibitem[\protect\citeauthoryear{{Walsh}, {Morbidelli}, {Raymond}, {O'Brien}
  \& {Mandell}}{{Walsh} et~al.}{2011}]{Walsh2011}
{Walsh} K.~J.,  {Morbidelli} A.,  {Raymond} S.~N.,  {O'Brien} D.~P.,
  {Mandell} A.~M.,  2011, \mn@doi [\nat] {10.1038/nature10201}, \href
  {http://adsabs.harvard.edu/abs/2011Natur.475..206W} {475, 206}

\makeatother
\end{thebibliography}
\bibliographystyle{mnras}
\bsp 
 
\label{lastpage}
\end{document}